\documentclass[aps,a4paper,showpacs,preprintnumbers,amsmath,amssymb,
 floatfix]{revtex4}
\usepackage{dcolumn}
\usepackage{bm}
\usepackage{amsmath,amssymb}
\usepackage{booktabs,subfigure}
\usepackage{graphicx,psfrag}
\usepackage{epsfig}
\usepackage{color} 
\usepackage{rotating}
\usepackage{axodraw}
\usepackage{slashed}
\usepackage{geometry}
\usepackage{rotating}
\allowdisplaybreaks[3]
\newcommand{\cw}[1][{}]{\ensuremath{\cos^{#1} \theta_{w}}}
\newcommand{\sw}[1][{}]{\ensuremath{\sin^{#1} \theta_{w}}}
\newcommand{\tw}[1][{}]{\ensuremath{\tan^{#1} \theta_{w}}}

\newcommand{\M}{\ensuremath{\mathcal{M}}}

\numberwithin{equation}{section} 
\def\lsim{\raise0.3ex\hbox{$\;<$\kern-0.75em\raise-1.1ex\hbox{$\sim\;$}}}
\def\gsim{\raise0.3ex\hbox{$\;>$\kern-0.75em\raise-1.1ex\hbox{$\sim\;$}}}
\begin{document}
\begin{flushright}
IMSc-2007/11/14
\end{flushright}
\title{
Radiative Neutralino Production in Low Energy 
Supersymmetric Models}

\author{Rahul Basu$^1$\footnote{\tt{Electronic address: rahul@imsc.res.in}}, 
P. N. Pandita$^{2,~3}$\footnote{\tt{Electronic address: ppandita@nehu.ac.in}} 
and Chandradew Sharma$^1$\footnote{\tt{Electronic address: sharma@imsc.res.in}}}
\affiliation{$^1$The Institute of Mathematical Sciences, Chennai 600 113, India}
\affiliation{$^2$ Department of Physics, North Eastern Hill University,
Shillong 793 022, India\footnote{permanent address}}
\affiliation{$^3$The Abdus Salam International Centre for Theoretical Physics,
Strada Costiera 11, 34014 Trieste, Italy}
%

\thispagestyle{myheadings}

 
\begin{abstract}
\noindent
We study the production of the lightest neutralinos in the radiative process
$e^+e^- \to \tilde\chi^0_1 \tilde\chi^0_1\gamma$ in low energy 
supersymmetric models for the International Linear Collider energies.  
This includes the minimal supersymmetric
standard model as well as its extension with an additional chiral
Higgs singlet superfield, the nonminimal supersymmetric standard model.
We  compare and contrast the dependence of the signal cross
section on the parameters of the neutralino sector of the minimal and 
nonminimal supersymmetric standard model.
We also consider the background to this process coming from the
Standard Model process  $e^+e^- \to \nu \bar\nu \gamma$, as well
as from the radiative production of the scalar partners of the neutrinos~(sneutrinos)
$e^+e^- \to \tilde\nu \tilde\nu^\ast \gamma$, which can be a background 
to the radiative neutralino production when the sneutrinos decay invisibly.
In low energy supersymmetric models
radiative production of the lightest neutralinos
may be  the only channel to study 
supersymmetric partners of the Standard Model particles at the
first stage of a linear collider, 
since heavier neutralinos, charginos 
and sleptons may be too heavy to be pair-produced at a $e^+ e^-$  
machine with $\sqrt{s} =500$ GeV.

\end{abstract}
\maketitle

\section{Introduction}
\label{sec:intro}
Supersymmetry~(SUSY) is at present the only  known framework~\cite{wess}
in which the Higgs sector of the Standard Model~(SM), so essential
for its internal consistency, is  technically natural~\cite{kaul}.
Supersymmetry is, however,  not an exact symmetry in nature.
The precise manner in which SUSY is broken is not known  at present.
However, the necessary SUSY breaking can be introduced through soft
supersymmetry breaking terms that do not reintroduce quadratic divergences in the 
Higgs mass, and thereby do not disturb the stability of the hierarchy between the
weak scale and the  large grand unified~(GUT) scale. Such terms can typically
arise in  supergravity theories, in which  local supersymmetry is spontaneously broken 
in a hidden sector, and is then transmitted to the visible sector via gravitational 
interactions. A particularly attractive implementation of the idea of supersymmetry, 
with soft supersymmetry breaking terms generated by supergravity, is the Minimal 
Supersymmetric Standard Model~(MSSM) obtained by simply introducing the supersymmetric 
partners of the  of the SM states, and introducing an additional Higgs doublet,
with opposite hypercharge to that of the SM Higgs doublet, in order to cancel the gauge 
anomalies and generate masses for all the fermions of the Standard 
Model~\cite{Nilles:1983ge, Drees:2004jm}. In order for broken supersymmetry to be 
effective in protecting the weak scale against large radiative corrections, 
the supersymmetric partners of the SM particles should have masses  of the order of
a few hundred GeV.  Their  discovery is one of the main goals of present and future 
accelerators.  In particular, a  $e^+e^-$ linear collider 
with a high luminosity ${\mathcal L}=500 fb^{-1}$,  and a center-of-mass
energy of $\sqrt s = 500$ GeV in the first stage,  will be
an important tool in determining the parameters of the
low energy supersymmetric model with a high 
precision~\cite{Aguilar-Saavedra:2001rg,Abe:2001nn,Abe:2001gc, 
Weiglein:2004hn,Aguilar-Saavedra:2005pw}.  Furthermore, polarisation
of the electron (and positron) beam can enhance the 
capability of such a linear collider~\cite{Moortgat-Pick:2005cw}
in unravelling the structure of the underlying supersymmetric
model.

In the minimal supersymmetric standard model the fermionic partners
of the two Higgs doublets~($H_1, H_2$) mix with the fermionic partners
of the gauge bosons to produce four neutralino states $\tilde \chi^0_i$,
$i = 1, 2, 3, 4$, and two chargino states  $\tilde \chi^{\pm}_j$,
$j = 1, 2.$  In the MSSM with $R$-parity~($R_p$) conservation, the lightest 
neutralino state is expected to be the lightest supersymmetric
particle~(LSP). The neutralino states of the minimal supersymmetric
standard model with $R_p$ conservation have been studied in great detail, because
the lightest neutralino, being the LSP, is the end product 
of any process involving supersymmetric particle in the final state.

However, the MSSM suffers from the so-called $\mu$ problem associated
with the bilinear term connecting the two Higgs doublet superfields
$H_1$ and $H_2$ in the superpotential.  An elegant solution to this
problem is to postulate the existence of a chiral electroweak gauge
singlet superfield $S$, and couple it to the two Higgs doublet
superfields $H_1$ and $H_2$ via a dimensionless trilinear term
$\lambda H_1 H_2 S$ in the superpotential. When the scalar component
of the singlet superfield $S$ obtains a vacuum expectation value, a
bilinear term $\lambda H_1 H_2 <S>$ involving the two Higgs doublets
is naturally generated.  Furthermore, when this scalar component of
the chiral singlet superfield $S$ acquires a vacuum expectation value
of the order of the $SU(2)_L \times U(1)_Y$ breaking scale, it gives
rise to an effective value of 
$\mu$~($\mu_{eff} \equiv \lambda <S> = \lambda x$) of the
order of the electroweak scale.  However, the inclusion of the singlet
superfield leads to additional trilinear superpotential coupling
$(\kappa/ 3 ) S^3$ in the model, the so called nonminimal, or
next-to-minimal~\cite{fayet, ellis89, drees89, nmssm1, nmssm2, nmssm3, nmssm4},
supersymmetric standard model~(NMSSM).  The absence of $H_1 H_2$ term,
and the absence of tadpole and mass couplings, $S$ and $ S^2$ in the
NMSSM is made natural by postulating a suitable discrete 
symmetry~\cite{Chemtob:2006ur, Chemtob:2007rg}.
The NMSSM is attractive on account of the simple resolution it 
offers to the $\mu$ problem, and of the scale invariance of its 
classical action in the supersymmetric limit. Since no dimensional 
supersymmetric parameters are present in the superpotential of NMSSM, 
it is the simplest supersymmetric extension of the Standard Model in 
which the electroweak scale originates from the supersymmetry breaking 
scale only.  Its enlarged Higgs sector may help in relaxing the 
fine-tuning and little hierarchy problems of the MSSM~\cite{nmssmft}, 
thereby opening new perspectives for the Higgs boson searches at high 
energy colliders~\cite{ellwang04,  moort05}, and for dark
matter searches~\cite{gunion05}. In the nonminimal supersymmetric standard 
model the mixing of fermionic partners of Higgs and gauge 
bosons~\cite{pnp1, pnp2, choi04}  produces five neutralino states  $\tilde \chi^0_i$,
$i = 1, 2, 3, 4, 5$, and two chargino states  $\tilde \chi^{\pm}_j$,
$j = 1, 2.$  Furthermore, because of the presence of the
fermionic partner of the singlet Higgs boson, the neutralino states can 
have an admixture of this $SU(2)_L \times U(1)_Y$ singlet fermion, 
thereby affecting the phenomenology of the
neutralinos in the nonminimal supersymmetric standard model.

The lightest neutralino state~($\tilde\chi_1^0$) of MSSM or NMSSM, being 
typically the LSP,  is 
stable and therefore, a possible dark matter 
candidate~\cite{Goldberg:1983nd, Ellis:1983ew}.
Since the neutralinos are among the lightest particles
in low energy supersymmetric models, 
they are expected to be the first states to
be produced at the colliding beam experiments.  
At an electron-positron collider, such as the  International
Linear Collider~(ILC), they can be directly produced in pairs
\begin{equation}
e^+ + e^-\to\tilde\chi_i^0 + \tilde\chi_j^0\,,
\label{pairneuts}
\end{equation}
which proceeds via $Z$ boson and selectron exchange~\cite{Bartl:1986hp, Ellis}.
In collider experiments the LSP
escapes detection such that the direct production of the lightest
neutralino pair
\begin{equation}
e^+ + e^-\to\tilde\chi_1^0 + \tilde\chi_1^0,
\label{lightestneut}
\end{equation}
is invisible.  Therefore, one must look for the signature of 
neutralinos in the radiative process
\begin{equation}
e^+ + e^-\to\tilde\chi_1^0 + \tilde\chi_1^0 + \gamma
\label{radiative}
\end{equation}
where the final photon is radiated off of the incoming beams or the 
exchanged selectrons.  We note that this process is suppressed 
by the square of the electromagnetic coupling. However,  it might be the
first proceess where the lightest supersymmetric states could  be observed at 
colliders.  The signal of the radiative process (\ref{radiative})
is a single high energy photon with the missing energy  carried away 
by the neutralinos.  The process~(\ref{radiative}) has been
studied in detail in the minimal supersymmetric  
model~\cite{Fayet:1982ky,Ellis:1982zz, Grassie:1983kq,
Kobayashi:1984wu,Ware:1984kq,Bento:1985in, Chen:1987ux,Kon:1987gi,
Choi:1999bs, Datta:1996ur, Ambrosanio:1995it}. In these studies,
different approximations have been used in calculating the
cross section for~(\ref{radiative}), and  the focus has been on 
LEP energies and special neutralino mixing, especially the case 
where the neutralino is a pure photino~\cite{Fayet:1982ky,Ellis:1982zz,
Grassie:1983kq,Kobayashi:1984wu, Ware:1984kq,Bento:1985in,
Chen:1987ux,Kon:1987gi}.

More recently, calculations have been carried out in the context of
MSSM assuming general neutralino 
mixing~\cite{ Choi:1999bs, Datta:1996ur, Ambrosanio:1995it}.
Some of these studies underline the importance of 
longitudinal~\cite{ Choi:1999bs} 
and even transverse beam polarisations. On the other hand, the signature 
``photon plus missing energy'' has been studied
in detail by different LEP collaborations ALEPH~\cite{Heister:2002ut},
DELPHI~\cite{Abdallah:2003np}, L3~\cite{Achard:2003tx}, and
OPAL~\cite{Abbiendi:2002vz,Abbiendi:2000hh}.  We recall that
in the SM, the radiative neutralino process  $e^+e^- \to \nu \bar\nu \gamma$ 
is the leading process with this signature, for which
the cross section depends on the number $N_\nu$ of light neutrino
species~\cite{Gaemers:1978fe}. This signature has, thus, been used to measure 
$N_\nu$, which has been found to be consistent with three.  Furthermore, 
the LEP collaborations have found no deviations from the SM prediction,
and, therefore,  only bounds on the masses of supersymmetric particles have 
been set~\cite{Heister:2002ut,Abdallah:2003np,Achard:2003tx,Abbiendi:2000hh}.  
This process is also important in determining collider
bounds on a very light neutralino~\cite{lightneutralino}.  For a
review of the experimental situation, see Ref.~\cite{Gataullin:2003sy}.

Most of the theoretical studies on radiative neutralino
production in the literature have been carried out in the framework of the
minimal supersymmetric standard model. 
This includes calculations relevant to ILC with a high center-of-mass energy, 
high luminosity and logitudinally polarized beams, as well as study of the SM 
background from the radiative neutrino production 
\begin{equation}
e^+e^- \to  \nu + \bar\nu + \gamma, 
\label{radiativenu} 
\end{equation}
and the MSSM background from radiative sneutrino 
production  
\begin{equation}
e^+e^- \to \tilde\nu + \tilde\nu^\ast + \gamma.
\label{radiativesnu}
\end{equation}
It has been pointed out~\cite{Dreiner:2006sb} that the 
discovery potential of the ILC might be significantly
extended if both beams are polarized, especially if other SUSY states 
like heavier neutralino, chargino or
even slepton pairs are too heavy to be produced at the first stage of
the ILC at $\sqrt s = 500$~GeV.  

In this paper we shall consider the radiative process (\ref{radiative})
in the nonminimal supersymmetric standard model. As discussed above, 
the nonminimal supersymmetric standard model is an attractive alternative
to the MSSM, which solves the  $\mu$ problem of the minimal supersymmetric 
standard model, and in which the weak scale originates from the supersymmetry 
breaking scale only. Furthermore, the NMSSM has five neutralino states in its 
spectrum, and there is an 
admixture of a singlet state in the neutralino states, which may 
affect  the radiative neutralino production process (\ref{radiative}).
On the other hand, the background processes from the SM and supersymmetry are 
not affected by the spectrum of neutralino states.
We shall compare and contrast the signal for the radiative neutralino
process in the NMSSM with that in the MSSM,  and study in detail
the dependence of the cross sections on the parameters of the neutralino sector.

The plan of the paper is as follows. In Sec.~\ref{sec: signal} we discuss 
the cross section for the signal process (\ref{radiative}) in NMSSM, and 
compare and contrast it with the corresponding cross section in the MSSM. 
Here we recall the basic features of the neutralino mixing matrix in NMSSM and MSSM, 
and the couplings of the neutralinos relevant for our calculations, as well as 
the cross section and the phase space for the signal process. 
Here we also describe the cuts on the photon angle and energy that
are  used to regularise the infrared and collinear divergences in the tree
level cross section.  We then describe the typical
set of input parameters for the NMSSM that are used in our numerical
evaluation of the cross sections.  The set of parameters that we use are obtained 
by imposing various experimental and theoretical constraints on the parameter space of
NMSSM. On the other hand, for MSSM we use the typical benchmark
parameter set of the  Snowmass Points and Slopes~1a~( SPS~1a) 
scenario~\cite{Allanach:2002nj}, except when otherwise indicated. 
We analyse numerically the dependence of the
cross section on the parameters of the neutralino sector, and on the
selectron masses.  In Sec.~\ref{sec:backgrounds} we discuss the backgrounds  to 
the radiative neutralino production process (\ref{radiative}) from
the SM and supersymmetric processes.  Here we also define a statistical significance 
for measuring an excess of photons from radiative neutralino production over the 
backgrounds.  We summarise our results and conclusions in Sec.~\ref{sec:conclusions}.  

\section{Radiative Neutralino Production}
\label{sec: signal}
\subsection{Neutralino Mass Matrix, Lagrangian and Couplings}
In order to calculate the cross section for the 
radiative production of neutralinos
\begin{eqnarray}
e^-(p_1) + e^+(p_2) \rightarrow \tilde{\chi}_1^0(k_1) + \tilde{\chi}_1^0(k_2)
+ \gamma(q),
\label{radiative1}
\end{eqnarray}
where the symbols in the brackets denote the four momenta of the respective particles,
we need to compute the couplings of the neutralinos to electrons and 
the scalar partners of electrons, the selectrons.  These can be obtained from the 
neutralino mixing matrix. To obtain the neutralino mixing matrix for the MSSM, 
we recall that the neutralino mass matrix obtains contributions from 
part of the MSSM superpotential 
\begin{eqnarray}
W_{\mathrm{MSSM}} & = & \mu H_1 H_2,
\label{WMSSM}
\end{eqnarray}
where $H_1$ and $H_2$ are the two Higgs doublet chiral superfields,
and $\mu$ is the supersymmetric Higgs(ino) parameter. In addition
to the contribution from the superpotential, the neutralino mass  matrix
receives contributions from the interactions between gauge and  matter
multiplets, as well as contributions from the soft supersymmetry breaking 
masses for the gauginos. Including all these contributions, the neutralino mass 
matrix, in the bino, wino, higgsino basis $(-i\lambda', -i\lambda^3, \psi_{H_1}^1,
\psi_{H_2}^2)$ can be written as~\cite{Bartl:1989ms, Haber:1984rc}
\begin{eqnarray}
\label{mssmneut}
M_{\mathrm{MSSM}} =
\begin{pmatrix}
M_1 & 0   & - m_Z \sw \cos\beta & \phantom{-}m_Z\sw \sin\beta \\
0   & M_2 & \phantom{-} m_Z \cw \cos\beta  & -m_Z \cw\sin\beta \\
 - m_Z \sw \cos\beta &\phantom{-} m_Z \cw \cos\beta  & 0 & -\mu\\
\phantom{-}m_Z\sw \sin\beta& -m_Z \cw\sin\beta & -\mu & 0
\end{pmatrix},
\end{eqnarray}
where $M_1$ and $M_2$ are the $U(1)_Y$ and the $SU(2)_L$
soft supersymmetry breaking gaugino mass parameters, respectively, and
$\tan\beta = v_2 /v_1$ is the ratio of the vacuum expectation
values of the neutral components of the two Higgs doublet 
fields $H_1$ and $H_2$, respectively. Furthermore,
$m_Z$ is the $Z$ boson mass, and $\theta_w$ is the
weak mixing angle. In the CP conserving case, $M$ is a real 
symmetric matrix and can be diagonalised by an orthogonal matrix. 
Since at least one
eigenvalue of $M$ is negative, we can use a unitary matrix $N$, the neutralino 
mixing matrix, to get a positive semidefinite diagonal matrix~\cite{Haber:1984rc} 
with the neutralino masses $m_{\chi_i^0}$: 
\begin{eqnarray}
\label{mssmdiag}
N^\ast M_{\mathrm{MSSM}} N^{-1} =   \mathrm{diag}\begin{pmatrix}m_{\chi_1^0}, 
& m_{\chi_2^0}, & m_{\chi_3^0}, & m_{\chi_4^0} \end{pmatrix}.
\end{eqnarray}
We note that that the transformation Eq.~(\ref{mssmdiag}) is only a
similarity transformation if $N$ is real.

For the NMSSM, the relevant part of the superpotential is
\begin{eqnarray}
\label{WNMSSM}
W_{\mathrm{NMSSM}} & = &  \lambda S H_1 H_2 - \frac{\kappa}{3}S^3,
\end{eqnarray}
where $S$ is the Higgs singlet chiral superfield.
In the basis   $(-i\lambda', -i\lambda^3, \psi_{H_1}^1,
\psi_{H_2}^2, \psi_S)$, the neutralino mass matrix for the NMSSM 
can then be written as~\cite{pnp1, pnp2}
\begin{eqnarray}
\label{nmssmneut}
M_{\mathrm{NMSSM}} = \begin{pmatrix}
M_1 & 0   & - m_Z \sw \cos\beta & \phantom{-}m_Z\sw \sin\beta & 0 \\
0   & M_2 & \phantom{-} m_Z \cw \cos\beta  & -m_Z \cw\sin\beta & 0 \\
- m_Z \sw \cos\beta &\phantom{-} m_Z \cw \cos\beta  & 0 & -\lambda x 
& -\lambda v_2\\
\phantom{-}m_Z\sw \sin\beta& -m_Z \cw\sin\beta & -\lambda x & 0 & -\lambda v_1\\
0 & 0 &  -\lambda v_2 & -\lambda v_1 & 2 \kappa x
\end{pmatrix},
\end{eqnarray}
where $<S> = x$ is the vacuum expectation value of the singlet Higgs field.
As in the case of MSSM, we can use a unitary matrix $N'$ to get
a positive semidefinite diagonal matrix with the neutralino masses
$m_{\chi_i^0}$~\cite{pnp1, pnp2}:
\begin{eqnarray}
\label{nmssmdiag}
N'^\ast M_{\mathrm{NMSSM}} N'^{-1} = 
\mathrm{diag}\begin{pmatrix}m_{\chi_1^0}, & m_{\chi_2^0}, & m_{\chi_3^0},
& m_{\chi_4^0} &  m_{\chi_5^0} \end{pmatrix}.
\end{eqnarray}
The Lagrangian for the interaction of neutralinos, electrons, selectrons and 
$Z$ bosons for MSSM is given by~\cite{Haber:1984rc}
\begin{eqnarray}
{\mathcal L} &=& (- \frac {\sqrt{2}e}{\cw} N_{11}^*)
                    \bar{f}_eP_L\tilde{\chi}^0_1\tilde{e}_R
                 + \frac{e}{\sqrt{2} \sw} (N_{12} + \tw N_{11}) 
                   \bar{f}_e P_R\tilde{\chi}^0_1\tilde{e}_L \nonumber \\
     & & + \frac{e}{4 \sw \cw} \left(|N_{13}|^2 - |N_{14}|^2\right)
           Z_\mu \bar{\tilde{\chi}}_1^0\gamma^\mu \gamma^5\tilde{\chi}_1^0
           \nonumber \\
          && + e Z_\mu \bar{f}_e \gamma^\mu 
             \big[ \frac{1}{\sw\cw}\left(\frac{1}{2} - \sw[2]\right) P_L 
                     - \tw  P_R\big] {f}_e + \mathrm{h. c.},
\label{mssmlagrangian}
\end{eqnarray} 
with the electron, selectron, neutralino and $Z$ boson fields denoted by
$f_e$, $\tilde{e}_{L,R}$, $\tilde{\chi}_1^0$, and $Z_\mu$, respectively,  
and $P_{R, L} = \frac{1}{2} \left(1 \pm \gamma^5\right)$. The corresponding
interaction Lagrangian for NMSSM is obtained from (\ref{mssmlagrangian})
by replacing $N_{ij}$ with  $N'_{ij}.$  The different  vertices 
following from (\ref{mssmlagrangian}) are shown in Table~\ref{feynmandiag}.
\begin{table}[h!]
\begin{center}
\caption{Vertices corresponding to various terms in the interaction 
Lagrangian (\ref{mssmlagrangian}) for MSSM. In addition we have 
also shown the vertices for selectron-photon  and electron-photon
interactions. The vertices for the NMSSM 
are obtained by replacing $N_{ij}$ with $N'_{ij}$.}
\vspace{5mm}
\begin{tabular}{lccccccccl}
\hline
\\
Vertex & & & & & & & & & Vertex Factor\\
& & &  & &&& &&\\
\hline
& & & & & && &&\\
right~selectron - electron - neutralino 
& & & & & && && {$\frac {-i e \sqrt{2}}{\cw}N_{11}^* P_L$}\\
& & & & & && &&\\
left~selectron - electron - neutralino
& & & & & && &&{$\frac{i e}{\sqrt{2} \sw} (N_{12} + \tw N_{11}) P_R$}\\
& & & & & && &&\\
neutralino - $Z^0$ - neutralino
& & & & & && && {$ \frac{i e}{4 \sw \cw} 
           \left(|N_{13}|^2 - |N_{14}|^2\right) \gamma^\mu \gamma^5$}\\
& & & & & && && \\
electron - $Z^0$ - electron
& & & & & && && {$ i e \gamma^\mu \big[ \frac{1}{\sw\cw}\left(\frac{1}{2} 
               - \sw[2]\right) P_L - \tw  P_R\big] $} \\
& & & & & && &&\\
selectron - photon - selectron
& & & & & && && {$ i e (p_1 + p_2)^\mu$}\\
& & & & & && && \\
electron - photon - electron
& & & & & && && {$ i e \gamma^\mu$}\\
& & & & &  && && \\
\\
\hline
\bottomrule
\end{tabular}
\label{feynmandiag}
\end{center}
\end{table}

The  couplings of the lightest neutralino to electrons, selectrons and $Z$ boson
are determined by the corresponding elements of the neutralino mixing 
matrix~($N_{ij}$  or $N'_{ij}$). For numerical calculation of the radiative 
neutralino cross section in the MSSM, we have chosen to work with the parameters  
in the SPS~1a scenario~\cite{Allanach:2002nj}. The parameters of the SPS~1a scenario are 
summarised in Table~\ref{parMSSM}.  However, since in the SPS~1a scenario the value 
of the parameters $\mu$ and $M_2$ are fixed, we shall use a different set of 
parameters to study the dependence of the neutralino mass and the radiative 
neutralino production cross section on $\mu$ and $M_2,$ and on the selectron masses. 
This set of parameters is shown in Table~\ref{parMSSMEWSB}. We shall call this set of 
parameters as the MSSM electroweak symmetry breaking scenario~(EWSB)~\cite{Pukhov:2004ca}.
\begin{table}[b!]
\renewcommand{\arraystretch}{1.0}
\caption{Input parameters and resulting  masses for various states in the MSSM SPS~1a scenario.}
\begin{center}
        \begin{tabular}{|c|c|c|c|}
\hline
$\tan\beta=10$ & $\mu= 358$~GeV & $M_2=192$~GeV & $m_0=100$~GeV\\
\hline \hline
$m_{\chi^0_{1}}=97$~GeV &
$m_{\chi^\pm_{1}}=180$~GeV &
$m_{\tilde e_{R}}=136$~GeV &
$m_{H}=400$ GeV \\
\hline
$m_{\chi^0_{2}}=181$~GeV &
$m_{\chi^\pm_{2}}=383$~GeV &
$m_{\tilde e_{L}}=196$~GeV &
$m_{h}=111$~GeV\\
\hline
\end{tabular}
\end{center}
\renewcommand{\arraystretch}{1.0}
\label{parMSSM}
\end{table}
\begin{table}[b!]
\renewcommand{\arraystretch}{1.0}
\caption{Input parameters and resulting masses of various states in  MSSM~EWSB scenario.}
\begin{center}
        \begin{tabular}{|c|c|c|c|}
\hline
$\tan\beta=10$ & $\mu= 130$~GeV & $M_1=150$~GeV & $M_2=300$~GeV\\
\hline
$M_3=1000$~GeV & $A_t=3000$~GeV & $A_b=3000$~GeV & $A_\tau=1000$~GeV\\
\hline \hline
$m_{\chi^0_{1}}=97$~GeV &
$m_{\chi^\pm_{1}}=119$~GeV &
$m_{\tilde e_{R}}=143$~GeV &
$m_{\tilde\nu_e}=194$ GeV \\
\hline
$m_{\chi^0_{2}}=-141$~GeV &
$m_{\chi^\pm_{2}}=330$~GeV &
$m_{\tilde e_{L}}=204$~GeV &
$m_{h}=120$~GeV\\
\hline
\end{tabular}
\end{center}
\renewcommand{\arraystretch}{1.0}
\label{parMSSMEWSB}
\end{table}
%
%
%
\begin{table}[h!]
\renewcommand{\arraystretch}{1.0}
\caption{Input parameters and resuling masses of various states in  NMSSM.}
\begin{center}
        \begin{tabular}{|c|c|c|c|}
\hline
$\tan\beta=10$ & $\mu= 130$~GeV & $M_1=150$~GeV & $M_2=300$~GeV\\
\hline
$M_3=1000$~GeV & $A_t=3000$~GeV & $A_b=3000$~GeV & $A_\tau=1000$~GeV\\
\hline
$\lambda = 0.55$ & $\kappa=0.44$ & $A_\lambda=880$~GeV &$A_\kappa= 10$~GeV \\
\hline \hline
$m_{\chi^0_{1}}=77$~GeV &
$m_{\chi^\pm_{1}}=121$~GeV &
$m_{\tilde e_{R}}=149$~GeV &
$m_{\tilde\nu_e}=194$ GeV \\
\hline
$m_{\chi^0_{2}}=-158$~GeV &
$m_{\chi^\pm_{2}}=334$~GeV &
$m_{\tilde e_{L}}=209$~GeV &
$m_{h}=122$~GeV\\
\hline
\end{tabular}
\end{center}
\renewcommand{\arraystretch}{1.0}
\label{parNMSSM}
\end{table}

On the other hand, for the NMSSM we use a set of parameters that is obtained
by imposing theoretical and experimental constraints on the parameter space of the NMSSM.
The parameters that enter the neutralino mass matrix of the NMSSM are, apart from
$M_1$ and $M_2$, $\tan\beta, \ \mu~(\equiv \lambda <S> = \lambda x), \  \lambda$ and
$\kappa$. For $M_1, M_2$ and $M_3$ we use the values which are 
consistent with the usual GUT relation $M_1/\alpha_1 = M_2/\alpha_2 = M_3/\alpha_3.$
We note that for the MSSM in  SPS~1a scenario, the value of the parameter 
$\tan\beta = 10 $. In order to remain close to the SPS~1a scenario of MSSM, 
we have chosen for our numerical calculations in NMSSM $\tan\beta = 10,$  whereas
the rest of the parameters  are chosen in such a way that the lightest Higgs boson mass, 
the lightest neutralino mass and the lightest chargino mass satisfy the present
experimental lower limits. We have also imposed on the parameter space of 
NMSSM the theoretical constraint that there is no charge and
color breaking global minimum of the scalar potential, and that a
Landau pole does not develop below the grand unified 
scale~($M_{GUT} \approx 10^{16}$ GeV).  For instance, for the parameter values 
$\tan\beta = 10 $, $\mu = 130$~GeV, $\lambda = 0.55 $ and $\kappa = 0.44 $,  we obtain
the lightest Higgs boson mass of about $123$ GeV, and the lightest
neutralino boson mass of about $77$ GeV, both of which are phenomenologically
acceptable values.  As another example, choosing  $\tan\beta = 10 $  and  
$\mu = 225$~GeV
results in the lightest Higgs boson mass of about $114$ GeV, the lightest
neutralino mass of about $132$~GeV but with  a charge and
color breaking global minimum. Other choices give a  Landau Pole
below $M_{GUT}$ etc. for some particular set of free parameters.
It is important to note that  choice of parameters away from
$\tan\beta = 10 $ is  also possible, for example, 
$\tan\beta = 3 $, $\mu = 180$~GeV, $\lambda = 0.3 $ and $\kappa = 0.1 $ 
produces a lightest Higgs boson mass of about $114$ GeV, and the lightest
neutralino mass of about $103$ GeV. We have taken $\tan\beta =  10 $ for NMSSM  
simply because we wish to compare our results with MSSM for the typical SPS~1a 
scenario.  The consequence  of imposing
these constraints on the parameter space of NMSSM, and the resulting  masses
for various particles for a particular choice of input parameters
is summarized in Table~\ref{parNMSSM}. 

\begin{figure}[h]
\setlength{\unitlength}{1cm}
{\scalebox{1}{\includegraphics{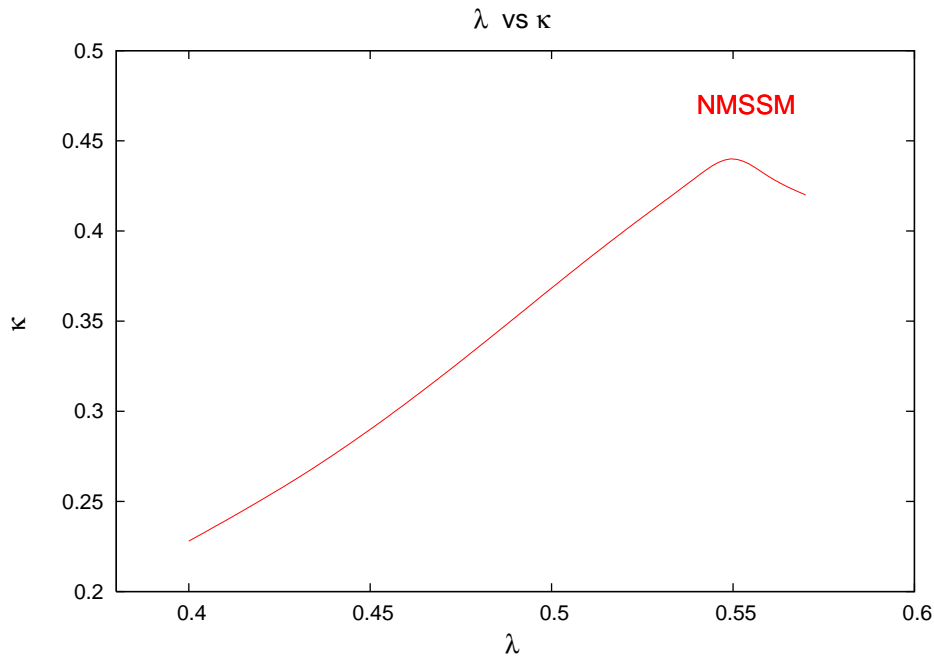}}}
\hspace{1mm}
\caption{Plot of $\lambda$ versus $\kappa$ for the set of input parameters
in  Table~\ref{parNMSSM}.}
\label{fig:lambda_kappa}
\end{figure}

Since the neutralino mass matrix depends on the parameters $\lambda$ and
$\kappa$, it is useful to study the possible values of these parameters, 
with all other parameters fixed, which satisfy the phenomenological
and theoretical constraints discussed above. In Fig.~\ref{fig:lambda_kappa}
we show a  plot of  $\lambda$ versus $\kappa$, with all other input
parameters fixed as in Table~\ref{parNMSSM}, and with the
lightest neutralino, the lightest Higgs boson, 
and the lightest chargino mass as in  Table~\ref{parNMSSM} with a
variation of
less than $5\%$.  Fig.~\ref{fig:lambda_kappa} shows the range of  
$\lambda$ and $\kappa$ values that are consistent with all the constraints 
discussed above for the set of input parameters in Table~\ref{parNMSSM}.

We note that for the set of input values in Table~\ref{parNMSSM}, values of
$\lambda \lsim 0.4$, with  $ \kappa \lsim 0.22$,
lead to an unphysical global minimum. On the other hand,
values of $\lambda \gsim 0.57$, with  $ \kappa \gsim 0.44$, 
lead to a Landau pole below the GUT scale.
Thus, the allowed values of $\lambda$ and $\kappa$, for the given set of 
input parameters, and for the fixed masses of lightest neutralino, 
the lightest Higgs boson, and the lightest chargino, as in Table~\ref{parNMSSM}
lie in a narrow range   $0.4 \lsim \lambda \lsim 0.57$ 
for $ 0.22 \lsim \kappa \lsim 0.44$.
For definiteness, we have chosen to work with the values of
$\lambda = 0.55$ and $\kappa = 0.44$ in this paper. These values correspond
to the peak in the $\lambda$ versus $\kappa$ plot in 
Fig.~\ref{fig:lambda_kappa}. 

\begin{figure}[htb]
\setlength{\unitlength}{1cm}
{\scalebox{1}{\includegraphics{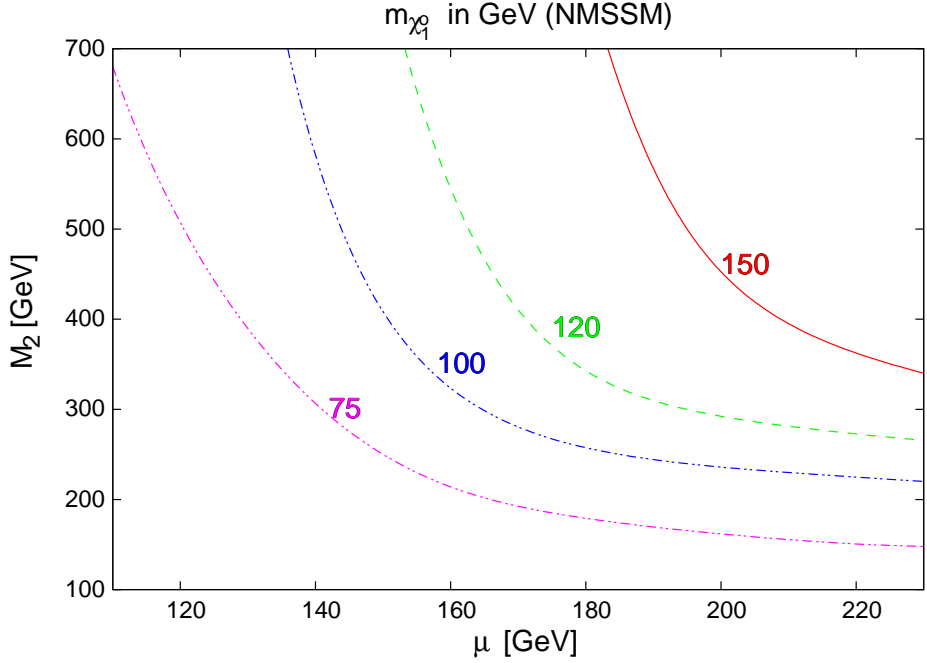}}}
\hspace{1mm}
\caption{Contour plots  of constant lightest  neutralino  mass $m_{\chi^0_1}$ in 
$\mu$ -$ M_2$ plane for NMSSM.}
\label{fig:muM2N}
\end{figure}

\begin{figure}[htb]
\setlength{\unitlength}{1cm}
{\scalebox{1}{\includegraphics{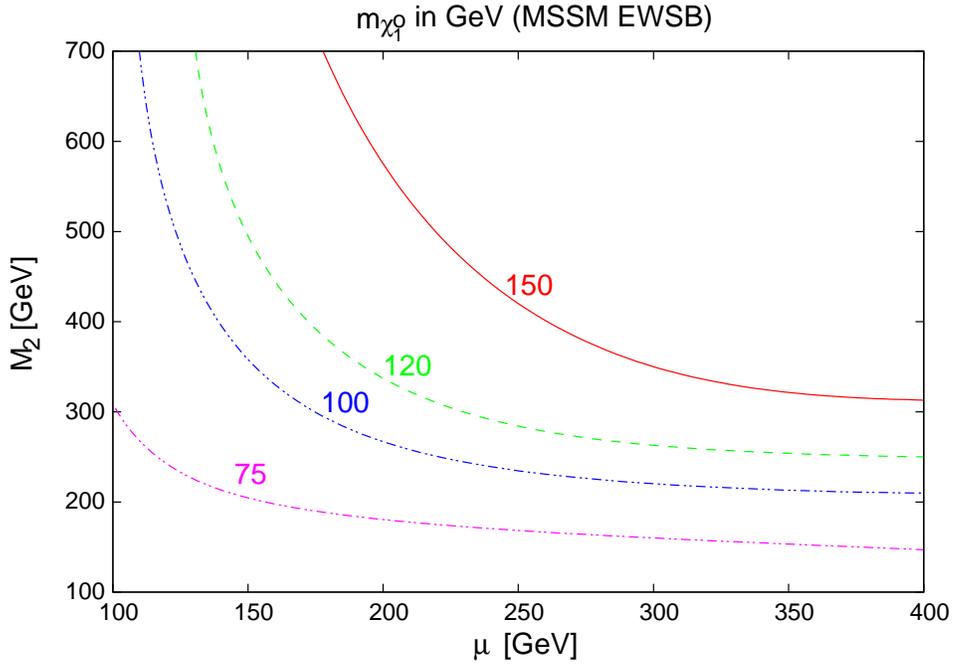}}}
\hspace{1mm}
\caption{Contour plots of constant lightest neutralino mass $m_{\chi^0_1}$ in 
  $\mu$ -$ M_2$ plane for  MSSM.}
\label{fig:muM2M}
\end{figure}

For the  parameters of Table~ \ref{parNMSSM}, the composition of the 
lightest neutralino in NMSSM is given by
\begin{eqnarray}
N'_{1j} & = & (0.39,~ -0.22,~ 0.57,~ -0.59,~  0.35).
\label{nmssmcomp}
\end{eqnarray}
{}From the composition (\ref{nmssmcomp}), we see that the lightest neutralino
has a sizable singlet component, thereby changing the  neutralino
phenomenology in the NMSSM as compared to MSSM. 
For comparison, we also show the particle content
of the lightest neutralino in MSSM
\begin{eqnarray}
N_{1j} & = & (0.5,~ -0.21,~ 0.66,~ -0.51)
\label{mssmcomp}
\end{eqnarray}
for the parameter set in Table~\ref{parMSSMEWSB}. In  Fig.~\ref{fig:muM2N} 
we have plotted the constant contour plots  for the mass
of lightest neutralino  in NMSSM in the $\mu$ - $M_2$ plane. We emphasize
that the choice of $\mu$ and  $M_2$ values in this plot  have been taken to be 
consistent with phenomenological and theoretical constraints as
described above. 
For comparison, we have also plotted the corresponding contour 
plots  for MSSM in
Fig.~\ref{fig:muM2M} with parameters as in Table ~\ref{parMSSMEWSB}. 

\subsection{Cross Section for the Signal Process}
\begin{figure}[h!]
{\unitlength=1.0 pt
\SetScale{1.0}
\SetWidth{1.0}      
\includegraphics{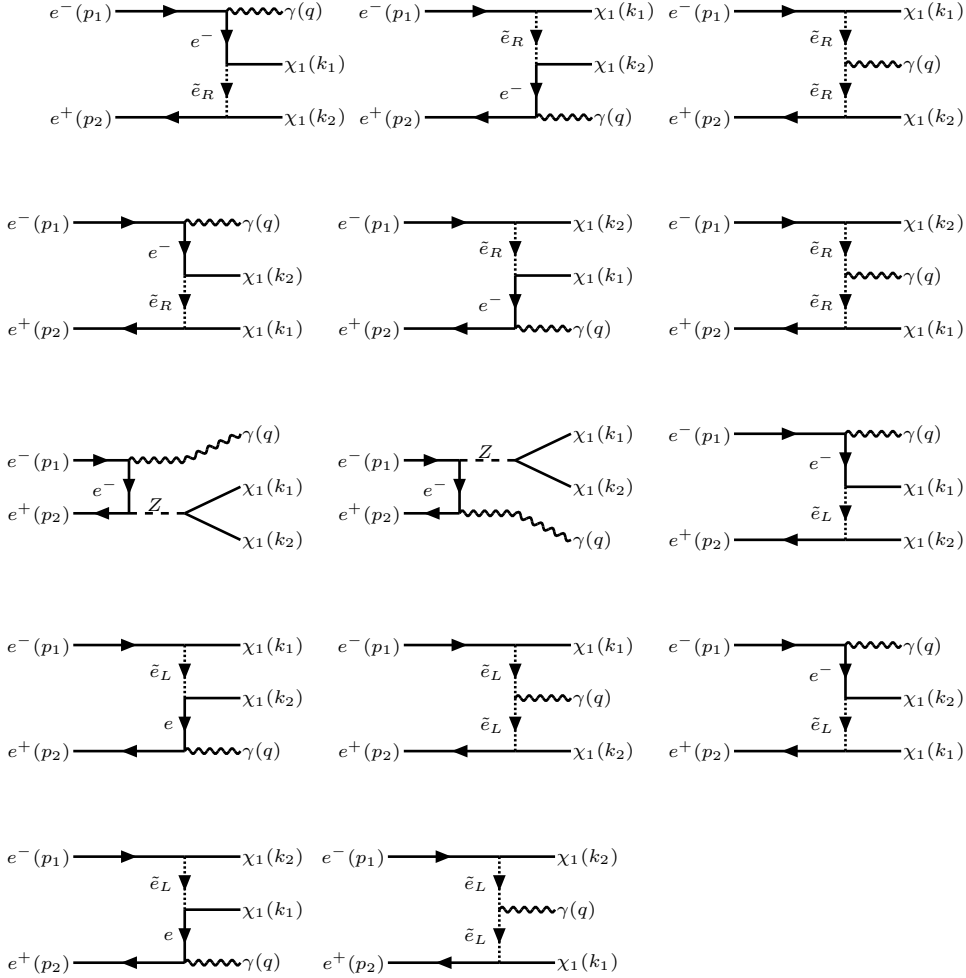}
}
\caption{Feynman diagrams contributing to the  radiative neutralino production 
 $e^+e^- \to  \tilde\chi_1^0\tilde\chi_1^0\gamma.$}
\label{fig:radneutralino}
\end{figure}

In NMSSM, and in MSSM, the process  (\ref{radiative1}) proceeds at the tree level
via $t$- and $u$-channel exchange of right and left selectrons
$\tilde e_{R,L},$  and via $Z$ boson exchange in the $s$ channel.
The corresponding Feynman diagrams are shown in Fig.~\ref{fig:radneutralino}.
The differential cross section for (\ref{radiative1}) can be written 
as ~\cite{Grassie:1983kq, Eidelman:2004wy}
\begin{eqnarray}
d \sigma &=& \frac{1}{2} \frac{(2\pi)^4}{2 s}
\prod_f \frac{d^3 \mathbf{p}_f}{(2\pi)^3 2E_f}\delta^{(4)}(p_1 +
p_2 - k_1 - k_2 - q)|\M|^2,
\label{phasespace}
\end{eqnarray}
where $\mathbf{p}_f$ and $E_f$ denote the final three-momenta
$\mathbf{k}_1$, $\mathbf{k}_2$, $\mathbf{q}$
and the final energies
$E_{\chi_1}$, $E_{\chi_2}$, and $E_\gamma$
of the neutralinos and the photon, respectively.
The squared matrix element $|\M|^2$ in (\ref{phasespace})
can be written as~\cite{Grassie:1983kq}
\begin{eqnarray}
|\M|^2 & = & \sum_{i \leq j} T_{ij}, \label{squaredmatrix}
\end{eqnarray}
where $T_{ij}$ are squared amplitudes corresponding to the Feynman diagrams
in  Fig.~\ref{fig:radneutralino}. An average over intial spins and a sum over the
spins of the outgoing neutralinos, as well as a sum over the polarizations
of the outgoing photon is included in $T_{ij}$. 
The phase space in (\ref{phasespace}) is described in~\cite{Grassie:1983kq}.
\noindent

\bigskip
\subsubsection{Numerical Results}
\label{sec:results}
The tree-level cross section for radiative neutralino production
(\ref{radiative1}), and the  background from radiative neutrino and sneutrino production,
(\ref{radiativenu}) and (\ref{radiativesnu}),  have been calculated
using the program CalcHEP~\cite{Pukhov:2004ca}. 
The tree level cross sections have infrared and collinear divergences, 
which need to be regularized. To do this we define the fraction of the
beam energy carried by the photon as $x = E_{\gamma}/E_{\rm beam},$
where  $ \sqrt {s} = 2E_{\rm beam}$ is the center of mass energy, and 
$E_{\gamma}$ is the energy carried away by the photon. 
We then impose the following cuts on $x$,  and on the scattering 
angle $\theta_{\gamma}$ of the 
photon~\cite{ Grassie:1983kq, Dreiner:2006sb}:
\begin{eqnarray}
0.02 \le x \le  1-\frac{m_{\chi_1^0}^2}{E_{\rm beam}^2}, \label{cut1} \\
\nonumber \\
-0.99 \le \cos\theta_\gamma \le 0.99.
\label{cut2}
\end{eqnarray}
The lower cut on $x$ in (\ref{cut1}) corresponds to a photon energy 
$E_\gamma= 5$\, GeV for $\sqrt{s}=500$\, GeV. The
upper cut of   $(1-m_{\chi_1^0}^2/E_{\rm beam}^2)$ on $x$  
corresponds to the kinematical limit of radiative neutralino production. 

In order to implement the cuts on the photon energy in the calculation
of the cross sections, we have taken the  mass of the lightest
neutralino in NMSSM to be $m_{\chi_1^0}=77$~GeV for the parameter 
set shown in Table~\ref{parNMSSM}.
For MSSM, we take $m_{\chi_1^0}=97$~GeV from the SPS1a scenario.
Using Eq.~(\ref{cut1}) we get
a fixed upper limit $E_\gamma^{\rm max}=226$~GeV for NMSSM
and $E_\gamma^{\rm max}=212$~GeV for MSSM at $\sqrt{s}=500$~GeV for the
photon energy. We have used exclusively these cuts for both signal and 
background processes. Two different mechanisms have
been chosen for the production of neutrinos in the background 
process~(\ref{radiativenu}), 
one  ``with upper cut''and another ``without upper cut,''  
for obvious reasons. 
We note that at $\sqrt{s}=500$ GeV and for $m_{\chi_1^0} \gsim 70$ GeV,
this cut reduces a substantial amount of the on-shell $Z$ boson 
contribution to radiative neutrino production process.         

\subsubsection{Photon Energy~($E_\gamma$) Distribution and Total Beam
Energy~($\sqrt {s}$) Dependence}
\begin{figure}[t!]
\setlength{\unitlength}{1cm}
{\scalebox{1}{\includegraphics{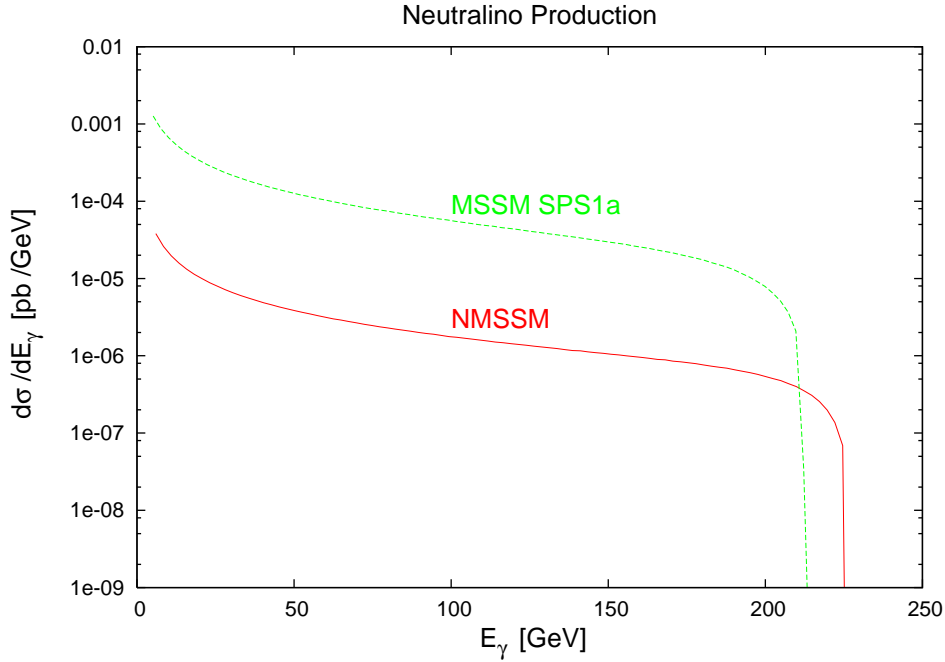}}}
\hspace{1mm}
\caption{ Photon energy
          distribution  {$\displaystyle \frac{d\sigma}{d E_\gamma}$}
          for the radiative neutralino production for
          NMSSM~(red solid line) and  for MSSM~(green dashed line) at $\sqrt{s} =
          500$ GeV.}
\label{fig:cdiffchi}
\end{figure}
\begin{figure}[t!]
\setlength{\unitlength}{1cm}
{\scalebox{1}{\includegraphics{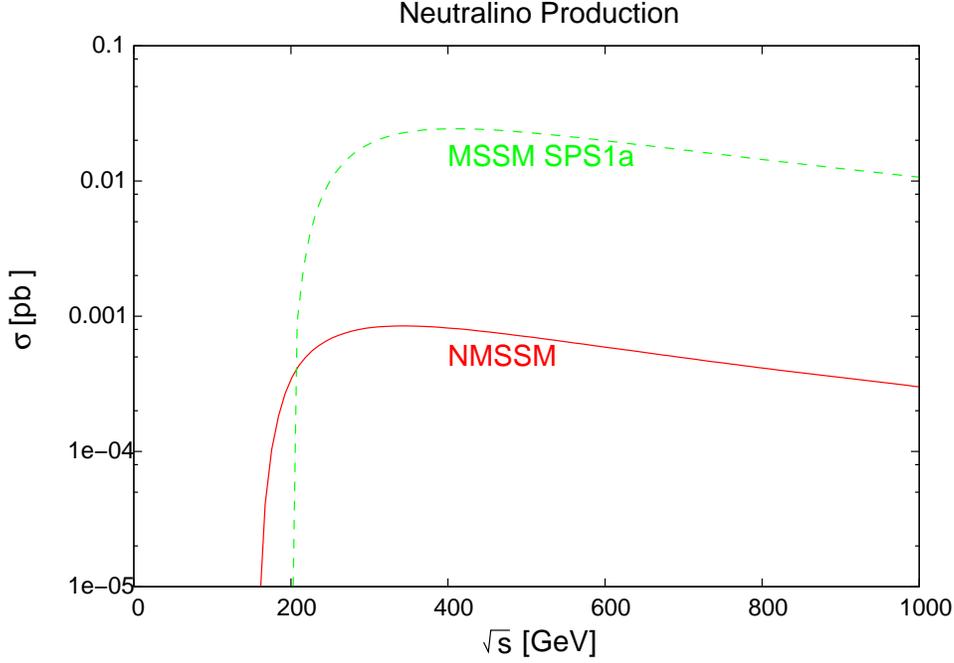}}}
\hspace{1mm}
\caption{ Total energy $\sqrt{s}$ dependence of
          the cross sections $\sigma$ 
         for radiative neutralino production $e^+e^- \to \tilde\chi^0_1
         \tilde\chi^0_1\gamma$ for NMSSM~(red solid line) and for MSSM  SPS1a 
         scenario~(green dashed line).}
\label{fig:ctotchi}
\end{figure}

Using the procedure described above, 
we have calculated  the energy distribution of the photons from
radiative neutralino  
production in  NMSSM and in MSSM SPS~1a scenario,  respectively.  This is shown in
Fig.~\ref{fig:cdiffchi},  where  we  compare  the
energy distribution of the photons in the two models. We note that 
the mass of the lightest neutralino is larger in MSSM~SPS1a scenario
than in NMSSM for the  set of parameters in  Tables~\ref{parMSSM}
and \ref{parNMSSM},  respectively, and consequently the upper 
cut for MSSM is lower than that for NMSSM.  We also show the total beam energy $\sqrt s$ 
dependence of the cross sections in Fig.~\ref{fig:ctotchi}  for NMSSM
and MSSM SPS~1a,  respectively. Due to a smaller value of the mass of
neutralino in  NMSSM compared to MSSM SPS1~a,  the total cross section in NMSSM is
less than that  in MSSM SPS~1a.  From the kinematical endpoint 
$E_\gamma^\mathrm{max} = E_\gamma^\mathrm{max}(m_{\chi^0_1})$  of the energy
distribution of the photon from radiative neutralino production,  the
neutralino mass can, in principle, be determined for $\sqrt{s}=500$~GeV.
We note that although the shape of the energy distribution and the total
cross section are similar in NMSS and MSSM, the numerical values for NMSSM
are considerably  smaller as compared to the MSSM. This is primarily due to the fact
that the lightest neutralino in NMSSM has a significant singlet component,
thereby reducing the cross section.
\begin{figure}[t!]
\setlength{\unitlength}{1cm}
{\scalebox{1}{\includegraphics{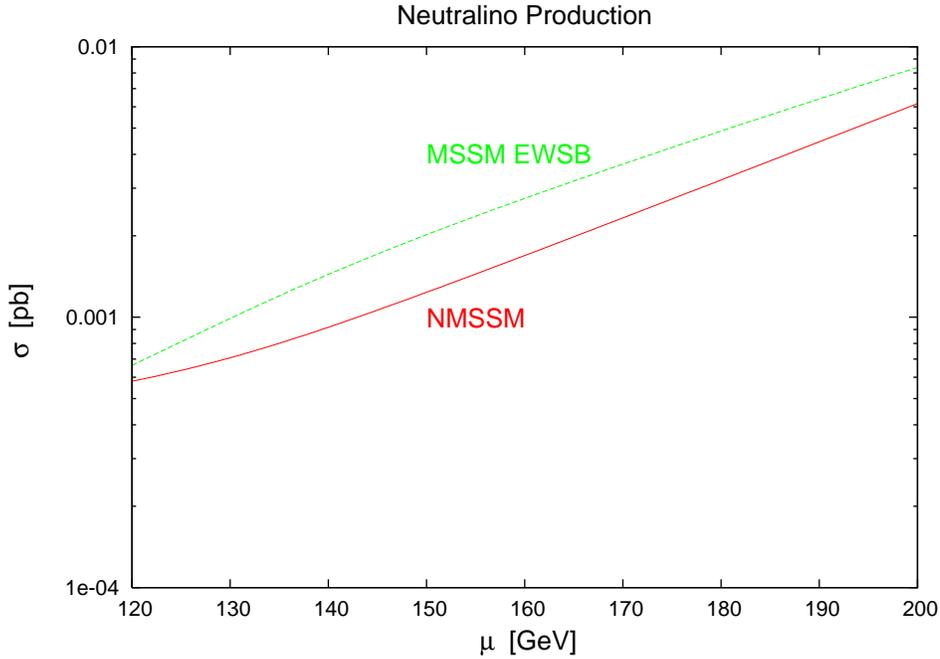}}}
\hspace{1mm}
\caption{Total cross-section $\sigma$ for  the radiative neutralino 
          production  versus $\mu$ for NMSSM~(red solid line) and for
          MSSM in the  EWSB scenario~(green dashed) at $ \sqrt {s} = 
          500 $ GeV.
	  For NMSSM $\mu \equiv \lambda <S> = \lambda x.$}
\label{fig:cmu}
\end{figure}
\begin{figure}[t!]
\setlength{\unitlength}{1cm}
{\scalebox{1}{\includegraphics{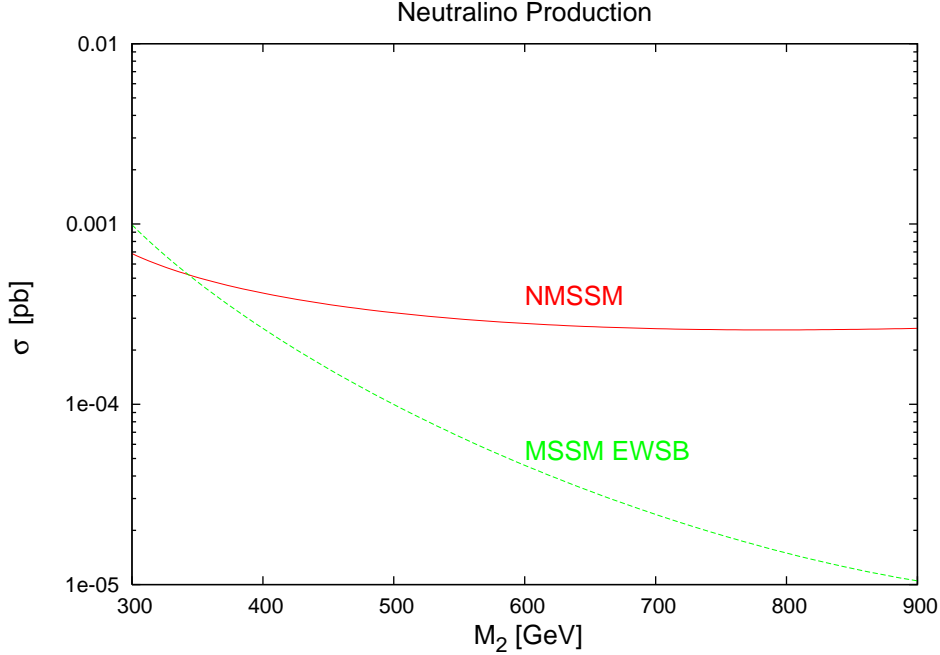}}}
\hspace{1mm}
\caption{ Total cross section $\sigma$ for  the radiative neutralino production 
versus $M_2$ for NMSSM~(red solid line) and for 
MSSM in the EWSB scenario~(green dashed) at $ \sqrt {s} = 500$ GeV.}
\label{fig:cM2}
\end{figure}
\subsubsection{Dependence on $\mu$ and $M_2$}
Since the neutralino mass matrix, and hence the lightest neutralino mass, depends on
$\mu$ and $M_2$, it is important  to study the dependence of the radiative neutralino
cross section on these parameters.
In the nonminimal supersymmetric standard model, $\mu~(\equiv \lambda <S> = \lambda x)$ 
and $M_2$ are independent parameters. We have, therefore, studied the  cross section 
$\sigma$($e^+e^- \to \tilde\chi^0_1 \tilde\chi^0_1\gamma$) as a function of
$\mu$ and  $M_2$ independently. 
In Fig.~\ref{fig:cmu}  we show the $\mu$ dependence
of the total cross section for the radiative production of neutralinos 
for  NMSSM as well as MSSM~(EWSB). We recall that in the MSSM SPS~1a
scenario these parameters are fixed. As is seen from Fig.~\ref{fig:cmu},
the total cross section increases with $\mu$. The plot  of total cross-section versus 
$\mu$ in Fig.~\ref{fig:cmu} is plotted in the range $\mu \in [120,200]$~GeV in
NMSSM and in MSSM~(EWSB). Note that the parameter values are chosen 
so as to avoid color and charge breaking minima, absence of Landau pole, and
the phenomenological constraints on different particle masses. 
Furthermore, in Fig.~\ref{fig:cM2},  we show the $M_2$ dependence of 
the total cross section for radiative neutralino production for NMSSM and MSSM.
The total cross-section decreases with increasing value of $M_2$. The graph of 
total cross-section versus $M_2$ in Fig.~\ref{fig:cM2} is plotted for  
the interval $M_2 \in [300,900]$~GeV in NMSSM and in MSSM~(EWSB) so as to 
satisfy the theoretical and phenomenoloigcal constraints described above. 
It is clear from Fig.~\ref{fig:cM2}
that the  radiative neutralino production cross section in  MSSM  decreases
sharply as compared to that in NMSSM.

\begin{figure}[t!]
\setlength{\unitlength}{1cm}
{\scalebox{1}{\includegraphics{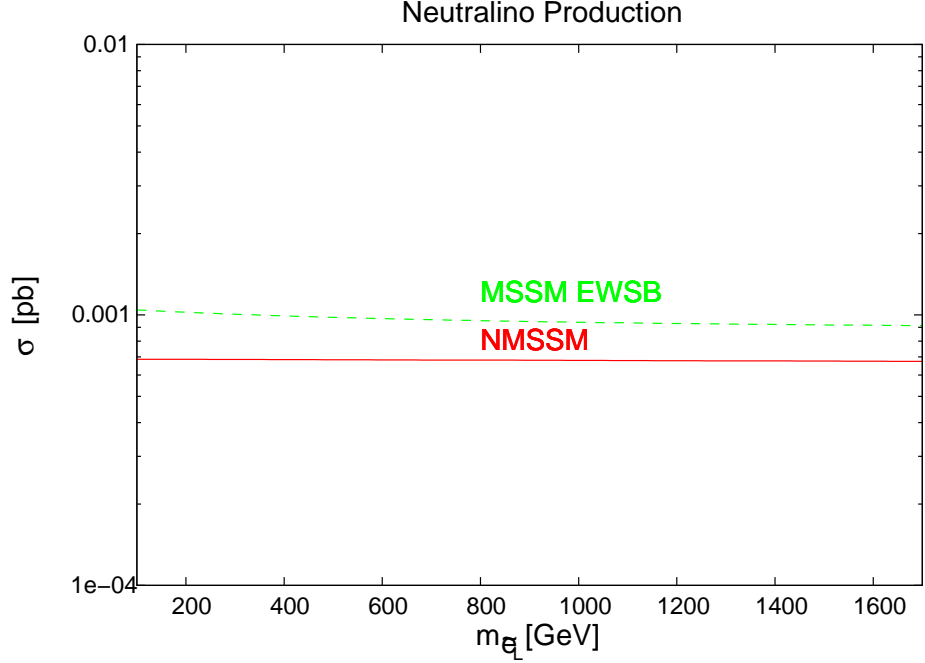}}}
\hspace{1mm}
\caption{ Total cross section $\sigma$ for  the radiative neutralino 
          production  versus $ m_{\tilde e_L}$ for NMSSM~(red solid line) and for
          MSSM in the  EWSB scenario~(green dashed) at 
        $ \sqrt {s} = 500$ GeV.}
\label{fig:smu}
\end{figure}
\begin{figure}[t!]
\setlength{\unitlength}{1cm}
{\scalebox{1}{\includegraphics{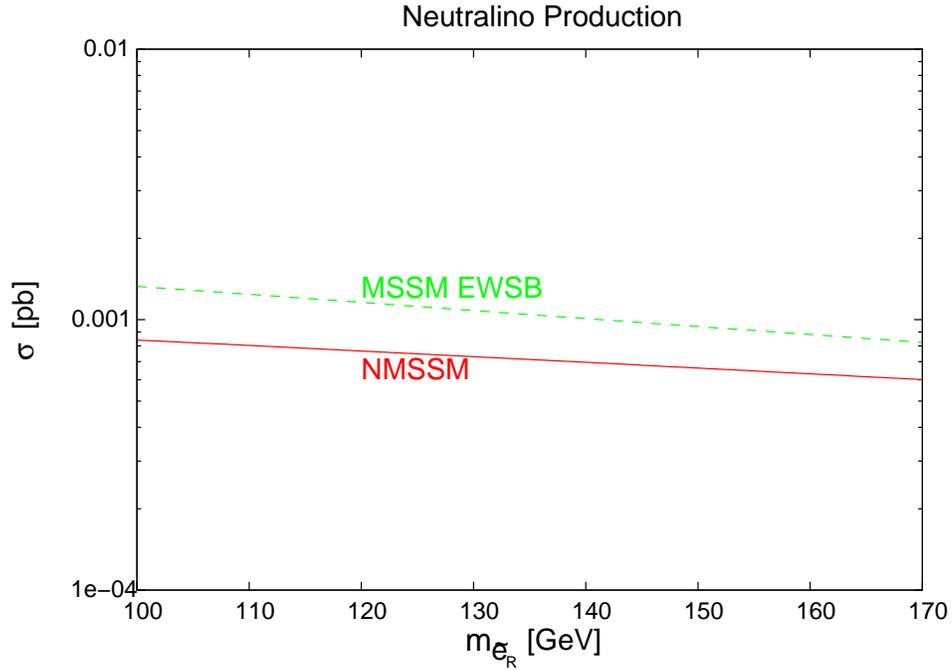}}}
\hspace{1mm}
\caption{ Total cross section $\sigma$ for  the radiative neutralino
  production  
versus $m_{\tilde e_R}$ for NMSSM~(red solid line) and for 
MSSM in the EWSB scenario~(green dashed) at $ \sqrt {s} = 500$ GeV.}
\label{fig:sM2}
\end{figure}
\subsubsection{Dependence on the Selectron Masses}
The cross section for radiative neutralino production $\sigma(e^+e^-
\to\tilde\chi^0_1\tilde\chi^0_1\gamma)$ proceeds mainly via right and left
selectron $\tilde e_{R,L}$ exchange in the $t$ and $u$-channels.  In the NMSSM
and MSSM~(EWSB),  the selectron masses are independent parameters. 
In Fig.~\ref{fig:smu} and Fig.~\ref{fig:sM2} we  show the 
dependence   of total cross section of radiative neutralino production 
on the left and right selectron masses. The cross section is not very sensitive to 
the selectron masses for both models. Furthermore, the total neutralino production 
cross section is smaller in NMSSM as compared to MSSM~(EWSB) as
a function of left as well as right selectron masses.  
\subsubsection{Photon Energy~($E_\gamma$) Distribution for the Radiative 
Production of the Second Lightest Neutralino}
\begin{figure}[t!]
\setlength{\unitlength}{1cm}
{\scalebox{1}{\includegraphics{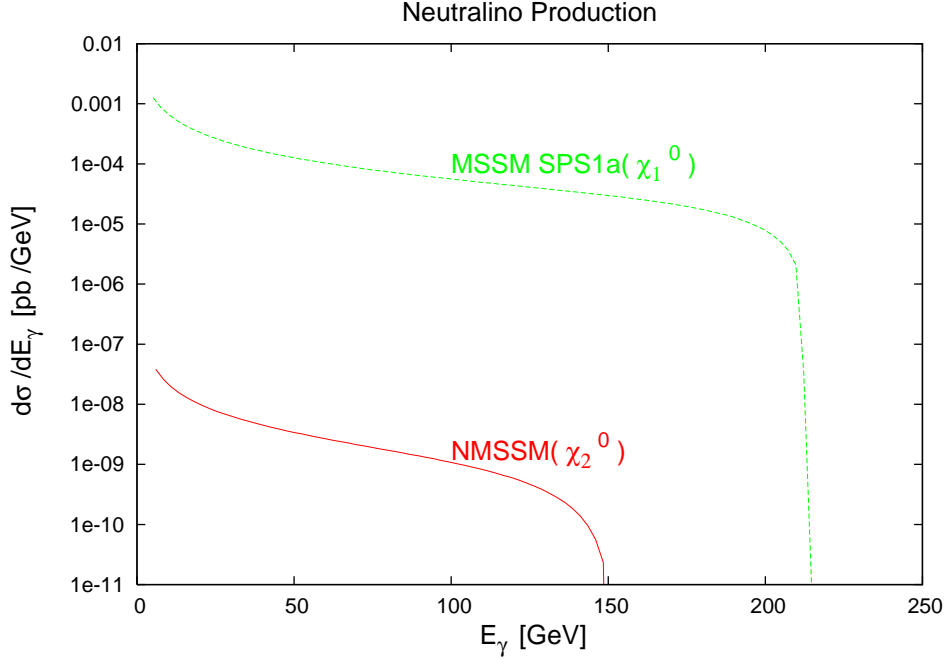}}}
\hspace{1mm}
\caption{ Photon energy
          distribution {$\displaystyle \frac{d\sigma}{d E_\gamma}$}
          for the radiative production of second lightest
          neutralino~($\chi_2^0$) for
          NMSSM~(red solid line) and  for the lightest neutralino~($\chi_1^0$)
          for  MSSM~(green dashed line) at $\sqrt{s} = 500$ GeV.}
\label{fig:cdiffchi2}
\end{figure}

As discussed above, the cross section for the production of the lightest
neutralino in NMSSM is rather small compared with the corresponding 
cross section for the lightest neutralino in the MSSM. It may,  therefore,
be useful to consider the radiative production of the second lightest 
neutralino in the NMSSM. For the parameter set of Table~\ref{parNMSSM} the
composition of the second lightest neutralino in NMSSM is given by
\begin{eqnarray}
N'_{2j} & = & (-0.085,~ 0.1,~ 0.69,~ 0.68,~ 0.19).
\label{nmssmcomp2}
\end{eqnarray}
We have calculated the photon energy distribution for the radiative 
production of the second lightest neutralino in NMSSM for the set of 
parameters shown in Table~\ref{parNMSSM}. This is shown in 
Fig.~\ref{fig:cdiffchi2}. For comparison we have also shown the photon 
energy distribution for the radiative production of the lightest 
neutralino in MSSM for the SPS~1a scenario. We see that the cross section 
for the production of the second lightest neutralino in NMSSM is much 
smaller than  the cross section for the lightest neutralino in MSSM.
\section {Background Processes}
\label{sec:backgrounds}
\subsection{The Neutrino Background}
\begin{figure}[htb]
\includegraphics{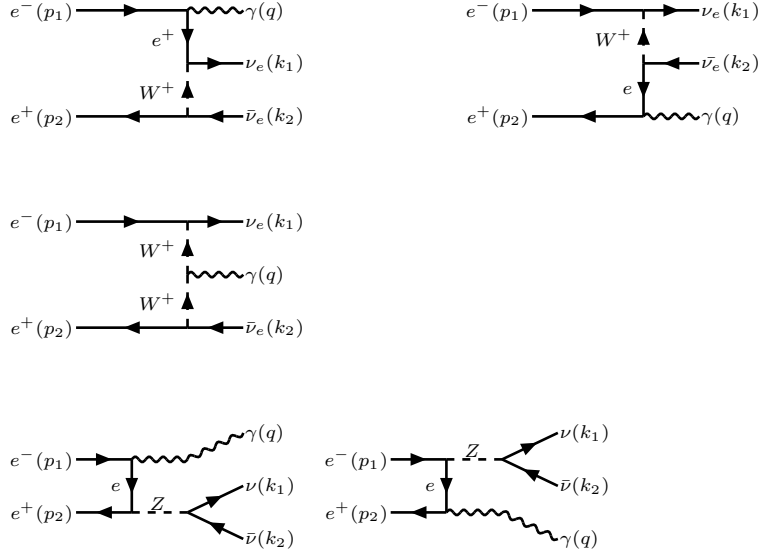}
\caption{Feynman diagrams contributing to the radiative neutrino process 
 $e^+e^- \rightarrow {\nu}{\bar{\nu}}\gamma.$}
\label{fig:radneutrino}
\end{figure}

\begin{figure}[htb]
\setlength{\unitlength}{1cm}
{\scalebox{1}{\includegraphics{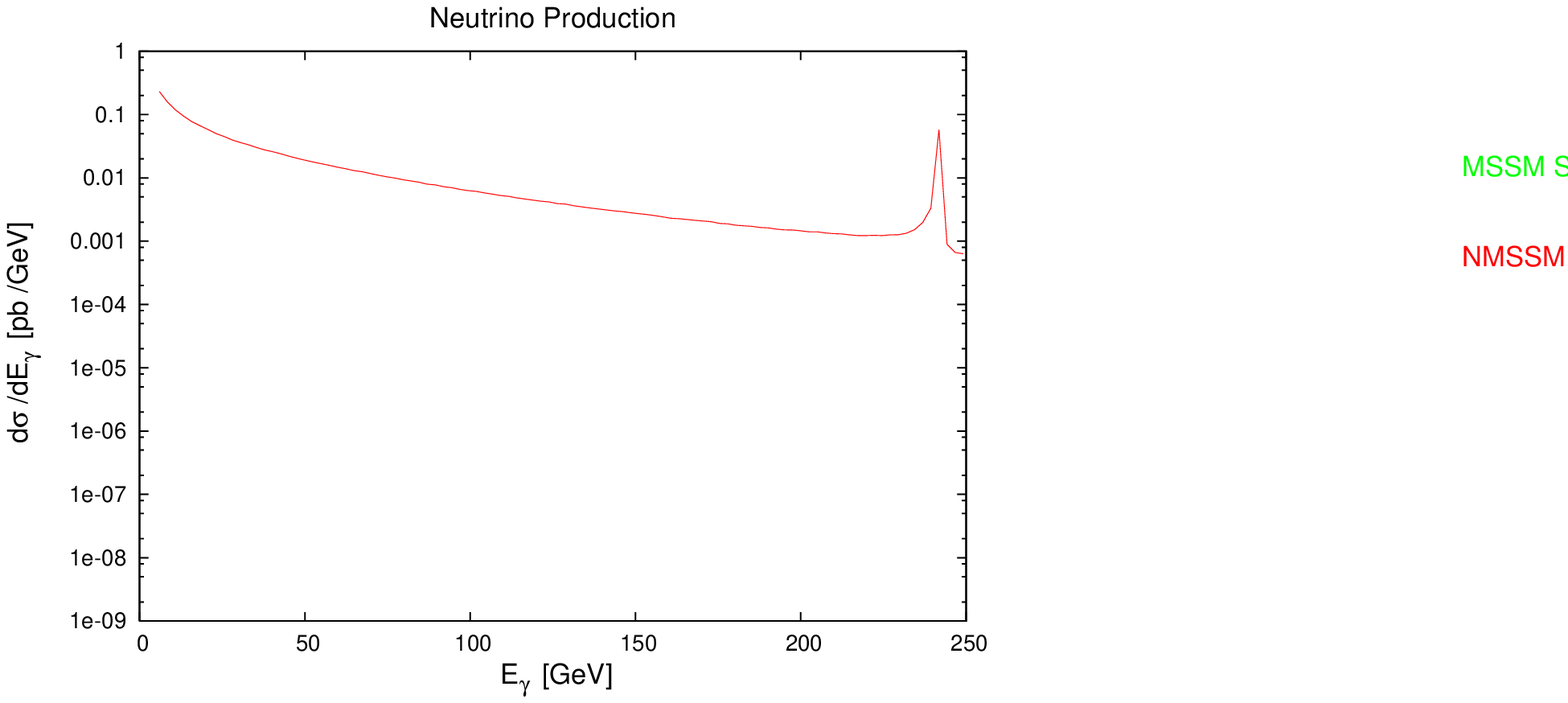}}}
\hspace{1mm}
\caption{ The photon energy distribution $\frac{d\sigma}{d E_\gamma}$
        for the radiative  neutrino  process  
        $e^+e^- \to \nu\bar\nu\gamma$  at $\sqrt {s} = 500$ GeV.}
\label{fig:ndiffneutrino}
\end{figure}

\noindent
The major background to the radiative neutralino production~(\ref{radiative1})
comes from the SM radiative neutrino production
process~\cite{Datta:1996ur,Gaemers:1978fe,Berends:1987zz,
Boudjema:1996qg,Montagna:1998ce}
\begin{equation}
e^+ +e^- \to \nu_\ell+\bar\nu_\ell+\gamma\,,\;\;\qquad \ell=e,\mu,\tau.
\label{radiative2}
\end{equation}
In this process $\nu_e$ are produced via
$t$-channel $W$ boson exchange, and $\nu_{e,\mu,\tau}$ via $s$-channel
$Z$ boson exchange.  We show the corresponding Feynman diagrams in
Fig.~\ref{fig:radneutrino}.  
\noindent
The background photon energy distribution $\frac{d\sigma}{d E_\gamma}$ and 
$\sqrt s$ dependence of the cross
section $\sigma$ for radiative  neutrino  production $e^+e^- \to
\nu\bar\nu\gamma$ is the  same for both NMSSM and MSSM.  
As shown in Fig.~\ref{fig:ndiffneutrino} the photon energy distribution from 
the radiative neutrino production peaks at
$E_\gamma= (s -m_Z^2)/(2\sqrt{s}) \approx 242$~GeV because of  
the radiative $Z$ production($\sqrt s > m_Z$). 
This  photon background from radiative neutrino production can be reduced  
by imposing an upper cut on the photon energy 
$x^{\rm max}=E_\gamma^{\rm max}/E_{\rm beam}=1-m_{\chi_1^0}^2
/E_{\rm beam}^2$, see Eq.~(\ref{cut1}), which is the kinematical
endpoint $E_\gamma^\mathrm{max}\approx 226$~GeV of the energy
distribution of the photon from radiative neutralino production
\begin{eqnarray}
 m_{\chi_1^0}^2 = \frac{1}{4}\left(s - 2\sqrt{s}E_\gamma^\mathrm{max}
\right).
\end{eqnarray}
In order to achieve this, one would have to separate the signal
and background processes. This would be possible if the neutralino is
heavy enough, such that the endpoint is  removed from the
$Z^0$ peak of the background distribution.

\begin{figure}[t!]
\setlength{\unitlength}{1cm}
{\scalebox{1}{\includegraphics{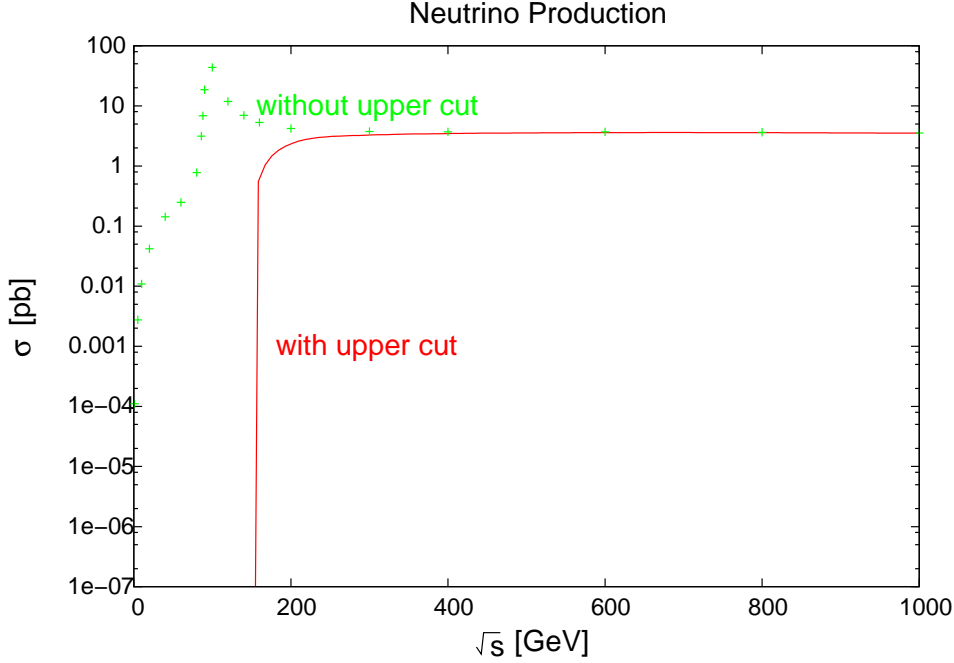}}}
\hspace{1mm}
\caption{ The total energy $ \sqrt {s}$
  dependence of the radiative  neutrino cross
  section $\sigma$($e^+e^- \to \nu\bar\nu\gamma$) with upper cut~(red line) on 
  the photon energy $E_\gamma,$ and of the radiative neutrino  cross 
  section without upper cut~(green  points) on the photon energy $E_\gamma$, 
  see Eq.~(\ref{cut1}).}
\label{fig:ntotneutrini}
\end{figure}

In Fig.~\ref{fig:ntotneutrini} we show the $\sqrt s$ dependence of the
total radiative neutrino cross section.  Without the upper cut on the photon energy 
$x^{\rm  max}$, the background cross section from
radiative neutrino production $e^+e^- \to \nu\bar\nu\gamma$~ 
(green  points in Fig.~\ref{fig:ntotneutrini}) is much larger than the
corresponding cross section with the cut~(solid red line).
However when we impose the cut, the signal cross section from radiative
neutralino production is  only about one order of magnitude smaller than the background
in the case of MSSM, but is nearly three orders of magnitude smaller than the background
in the case of NMSSM.
\bigskip

\subsection{The Supersymmetric Background}
\begin{figure}[htb]
{%
\unitlength=1.0pt
\includegraphics{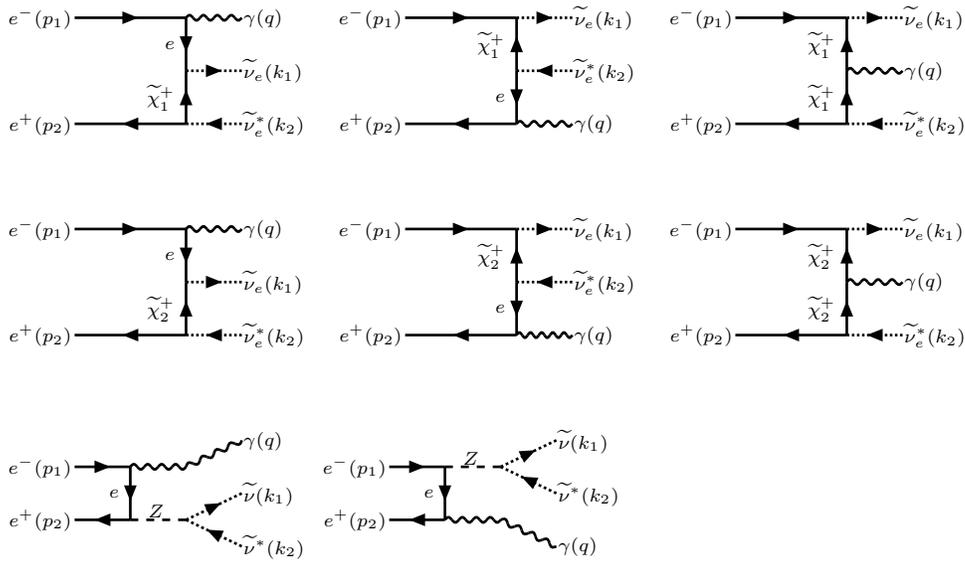}
}
\caption{Feynman diagrams contributing  to the radiative sneutrino production
process $e^+e^- \rightarrow \tilde{\nu}\tilde{\nu}^*\gamma.$}
\label{fig:radsneutrino}
\end{figure}

\begin{figure}[t!]
\setlength{\unitlength}{1cm}
{\scalebox{1}{\includegraphics{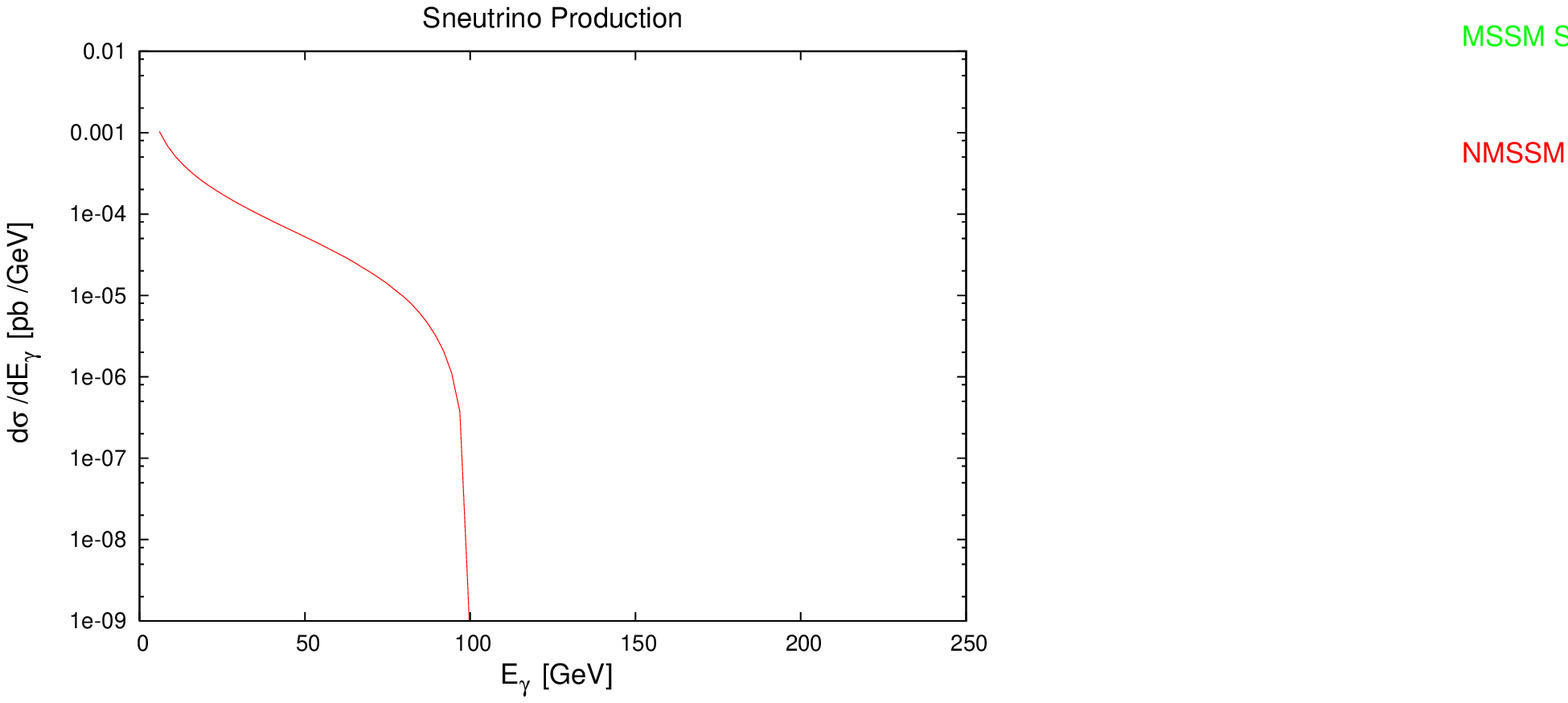}}}
\hspace{1mm}
\caption{ The photon energy distribution $\frac{d\sigma}{d E_\gamma}$
        for the radiative  sneutrino  production 
         $e^+e^- \to \tilde\nu\tilde\nu^\ast\gamma$  
         at $\sqrt{s} = 500$ GeV.}
\label{fig:ndiffsneutrino} 
\end{figure}

\begin{figure}[t!]
\setlength{\unitlength}{1cm}
{\scalebox{1}{\includegraphics{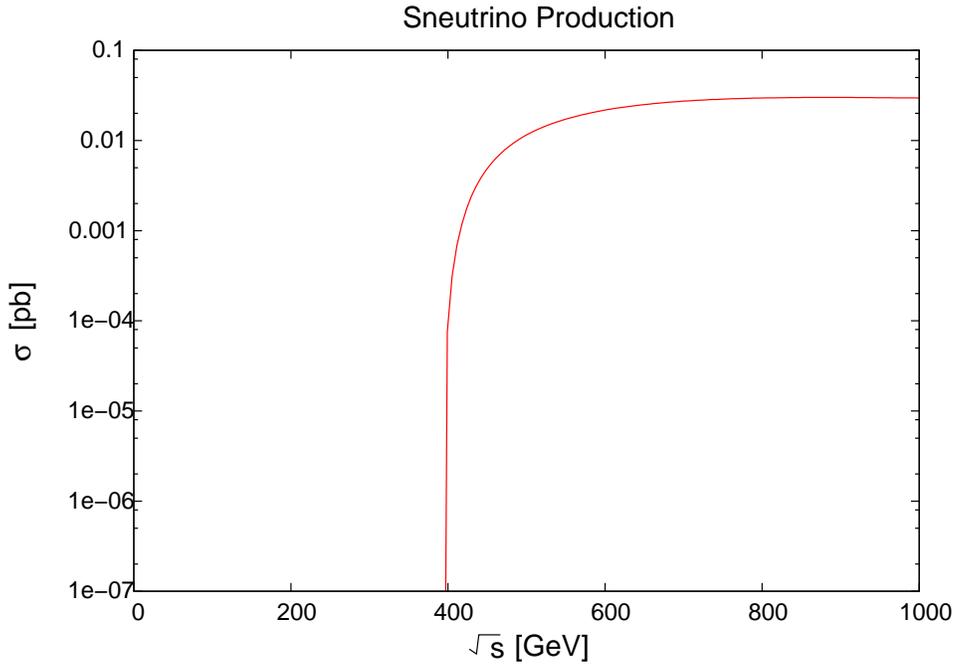}}}
\hspace{1mm}
\caption{ Total energy $\sqrt {s}$
  dependence of the radiative  sneutrino production cross
  section $\sigma$($e^+e^- \to \tilde\nu\tilde\nu^\ast\gamma$).}
\label{fig:ntotsneutrino}
\end{figure}

Apart from the SM background coming from (\ref{radiative2}), the radiative
neutralino production  (\ref{radiative1}) has a background coming from the
supersymmetric sneutrino production process~\cite{Datta:1996ur, Franke:1994ph}
\begin{equation}
e^+ +e^- \to \tilde\nu_\ell+\tilde\nu^\ast_\ell+\gamma\,, 
\;\qquad \ell=e,\mu,\tau\,.
\label{radiative3}
\end{equation}
The lowest order Feynman diagrams contributing to the process 
(\ref{radiative3}) are shown in Fig.~\ref{fig:radsneutrino}.
This process receives $t$-channel contributions via
virtual charginos for $\tilde\nu_e\tilde\nu_e^\ast $-production, as
well as $s$-channel contributions from $Z$ boson exchange for
$\tilde\nu_{e, \mu,\tau}\tilde\nu_{e, \mu,\tau}^\ast $-production. 
In Fig.~\ref{fig:ndiffsneutrino}, we show the photon energy distribution 
$\frac{d\sigma}{d E_\gamma}$ for radiative  sneutrino  production
$e^+e^- \to \tilde\nu\tilde\nu^\ast\gamma$ at $\sqrt{s} = 500$ GeV.
The total cross section for the radiative sneutrino production
is shown in Fig.~\ref{fig:ntotsneutrino}.

Radiative sneutrino production (\ref{radiative3}) can be a major supersymmetric 
background to neutralino production (\ref{radiative1}) if  sneutrinos
decay mainly invisibly, e.g.  via $\tilde\nu\to\tilde \chi^0_1\nu$.
This leads to so called ``virtual LSP'' scenario~\cite{Datta:1996ur}.  
However, if kinematically allowed, other visible decay channels 
like $\tilde\nu\to\tilde\chi^\pm_1\ell^\mp$ reduce the background rate 
from radiative sneutrino production. For example in the SPS~1a 
scenario~\cite{Ghodbane:2002kg, Allanach:2002nj} of the MSSM we have 
${\rm  BR}(\tilde\nu_e\to\tilde\chi_1^0\nu_e)=85\%$.

Furthermore,  neutralino production $e^+e^- \to \tilde\chi_1^0
\tilde\chi^0_2$ followed by subsequent radiative neutralino
decay~\cite{Haber:1988px} $\tilde\chi^0_2 \to \tilde\chi^0_1 \gamma$
is also a potential background.  However, significant branching ratios
${\rm BR}(\tilde\chi^0_2 \to \tilde\chi^0_1 \gamma)>10\%$ are only
obtained for small values of $\tan\beta<5$ and/or $M_1\sim
M_2$~\cite{Ambrosanio:1995it,Ambrosanio:1995az,Ambrosanio:1996gz}.
Thus,  we neglect this background, detailed  discussions
of which  can be found in Refs.~\cite{Ambrosanio:1995az,Ambrosanio:1996gz,Baer:2002kv}.
\noindent

\begin{figure}[t!]
\setlength{\unitlength}{1cm}
{\scalebox{1}{\includegraphics{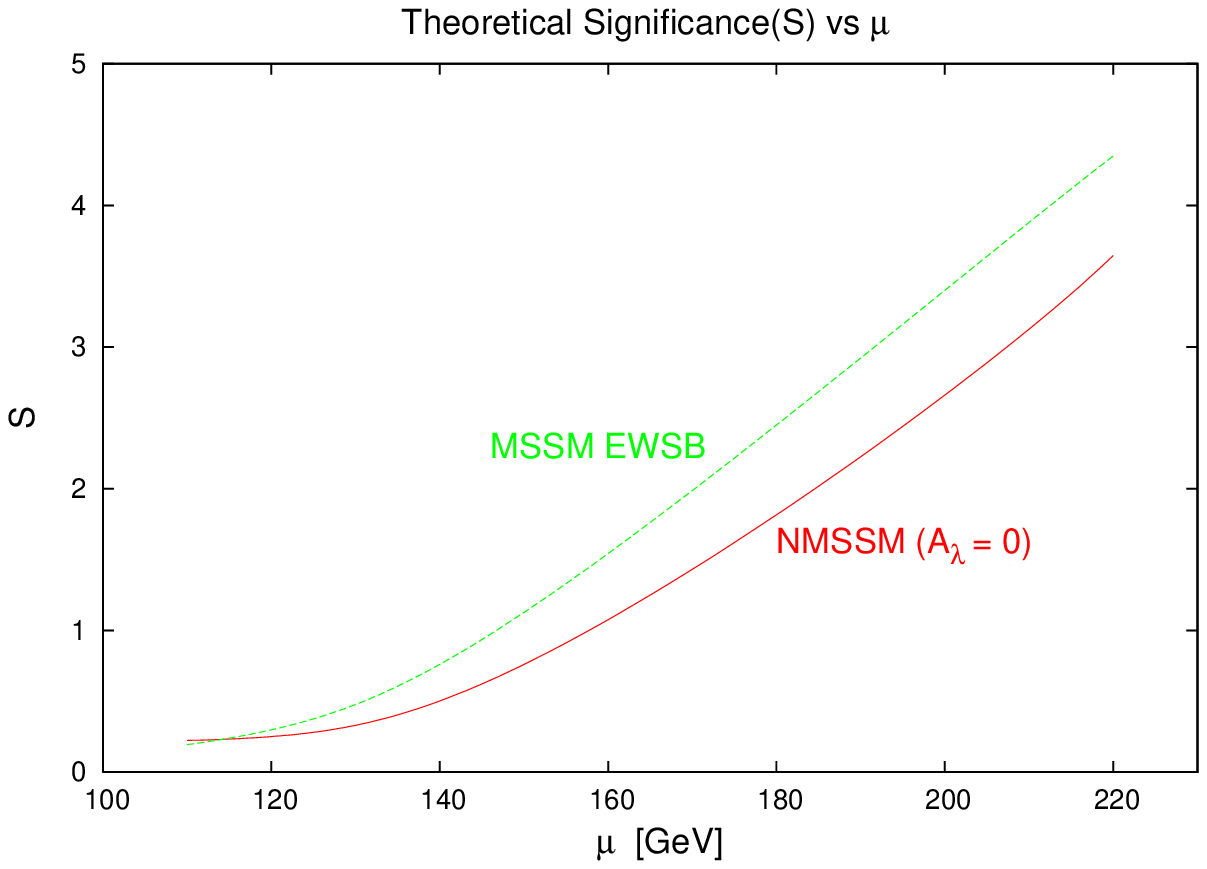}}}
\hspace{1mm}
\caption{ Theoretical significance
$S = \frac{\sigma}{\sqrt{\sigma + \sigma_{\rm B}}} \sqrt{\mathcal L}$
for  the radiative neutralino production  versus $\mu$
for NMSSM~(red solid line) and for MSSM in the
EWSB scenario~(green dashed). The value of $ \sqrt {s} = 500$~ GeV.}
\label{fig:tmu}
\end{figure}
\begin{figure}[t!]
{\scalebox{1}{\includegraphics{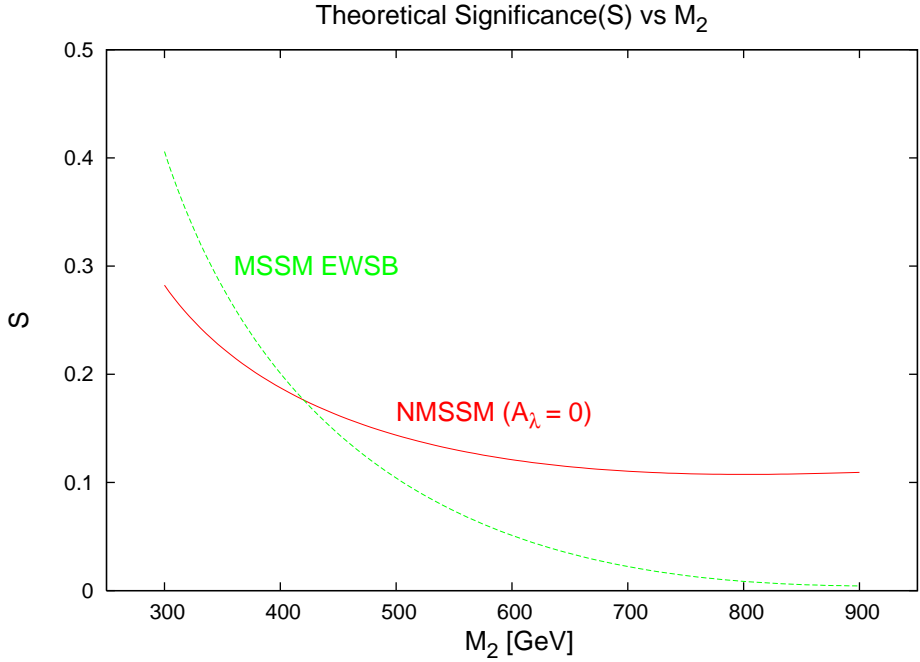}}}
\hspace{1mm}
\caption{ Theoretical significance
$S = \frac{\sigma}{\sqrt{\sigma + \sigma_{\rm B}}} \sqrt{\mathcal L}$
for  the radiative neutralino  production
versus $M_2$ for NMSSM~(red solid line) and for
MSSM in the EWSB scenario~(green dashed). The value of  $ \sqrt {s} = 500~$ GeV.}
\label{fig:tM2}
\end{figure}
\subsection{Theoretical Significance}
We now consider the question as to whether an excess of  photons from
radiative neutralino production can be measured over the SM background photons
coming from radiative neutrino production. To quantify the excess of photons
from the signal over the SM background photons for a given integrated
luminosity $\mathcal{L}$, 
we consider the theoretical significance~\cite{Dreiner:2006sb} 
\begin{equation}
S  =  \frac{N_{\rm S}}{\sqrt{N_{\rm S} + N_{\rm B}}}=
\frac{\sigma}{\sqrt{\sigma + \sigma_{\rm B}}} \sqrt{\mathcal L} ,
\label{significance}
\end{equation}
where $N_{\mathrm{S}}=\sigma {\mathcal L}$ is the number of signal photons,
and $N_{\rm B}=\sigma_{\rm B}{\mathcal L}$ is the number of photons
from the SM background of radiative neutrino production process.

The processes $e^+e^- \to \tilde\chi^0_1 \tilde\chi^0_1\gamma$
and $e^+e^- \to \nu \bar\nu \gamma$ depend significantly on the beam
energy only near threshold in most of the parameter space for 
$\sqrt{s}= 500$ GeV 
and $\mathcal L = 5 \times 10^5 $ pb$^{-1}$.
In Fig.~\ref{fig:tmu},  we show the $\mu$ dependence
of the theoretical significance $S$
for  NMSSM.
For this plot we have chosen $A_{\lambda} = 0$, and varied 
the parameter $\mu$ in the interval $\mu \in [110,220]$~GeV.
All other parameters for the NMSSM are chosen as in Table~\ref{parNMSSM}.
For comparison we have also also shown in Fig.~\ref{fig:tmu}
the theoretical significance  $S$ for the MSSM with parameters as
in Table~\ref{parMSSMEWSB} for the EWSB model.
We note that in NMSSM for  $\mu < 110$~GeV, and with other parameters as 
described above, the Landau pole 
develops below the GUT scale, and the lightest chargino and the lightest
neutralino masses are below the experimental lower bounds. 
On the other hand, for $\mu > 220$~GeV, the NMSSM develops 
unphysical global minima. We note from this analysis  that
Significance of $S\simeq 4$ can be attained for  $\mu \simeq 220$~GeV.

We have also studied the behavior of the theoretical significance
as a function of the $SU(2)_L$ gaugino mass $M_2$. In
Fig.~\ref{fig:tM2},  we show the $M_2$ dependence of
the  theoretical significance $S$ for  NMSSM and MSSM in the interval 
$M_2 \in [300,900]$~GeV. As in the study of the $\mu$ dependence, for
NMSSM, we have chosen  $A_{\lambda} = 0$, and all other parameters as 
in Table~\ref{parNMSSM}.  Furthermore, for the NMSSM the interval for 
$M_2$ is  chosen in order to satisfy the theoretical and experimental 
constraints. For $M_2 <300$~GeV, the Landau pole for NMSSM develops
below the GUT scale. On the other hand for values of $M_2 > 900$,
there is a sfermion with negative mass squared in the spectrum of
NMSSM. We note from Figs.~\ref{fig:tmu} and \ref{fig:tM2}
that higher values of $\mu$ and lower values of $M_2$ are favored for
achieving higher values of the significance $S$.
A theoretical significance of $S = 1$ would mean 
that the signal can be measured at a $68~\%$ confidence level.
On the other hand a significance of $5$ is required for the detection
of the signal. We note that a signal significance of about $4$ can be 
achieved for NMSSM for values of $\mu \simeq 220$~GeV.
Besides the theoretical significance, one must also consider the signal
to background ratio $N_S/N_B$ in order to judge the reliability of the 
analysis. It will be necessary to do a detailed Monte Carlo analyses to 
predict the significance.  However, this is beyond the scope of the 
present work. 
\section{Summary and Conclusions}
\label{sec:conclusions}

The nonminimal supersymmetric standard model is an attractive low energy
supersymmetric model which solves the $\mu$ problem of the minimal
supersymmetric standard model. We have carried out a detailed 
analysis of the radiative neutralino production $e^+e^- \to
\tilde\chi^0_1 \tilde\chi^0_1\gamma$  in the NMSSM for the International
Linear Collider energies and compared it with the corresponding results in the MSSM.
This process has a signature of a high energy photon and missing energy.
We have  obtained a typical set of parameter values for the NMSSM by imposing 
theoretical and experimental constraints on the parameter space of NMSSM.
For the  set of parameter values that we obtain in this manner, 
the lightest neutralino in NMSSM
has a significant admixture of the fermionic component of the singlet chiral
superfield $S$. Using this parameter set, we have studied in detail  the 
radiative neutralino production cross section in NMSSM for the ILC energies with
unpolarized $e^+$ and $e^-$ beams. For comparison, 
we have used the  SPS~1a benchmark scenario for the MSSM.
We have also calculated the background to this process 
from the SM  process  $e^+e^- \to \nu \bar\nu \gamma$,
as well as the background from the supersymmetric process
$e^+e^- \to \tilde\nu \tilde\nu^\ast \gamma$. All these processes have 
a signature of a highly energetic photon with missing energy.
The photon energy distribution $d\sigma/dE_{\gamma}$, 
and the total cross section as  a function of the total energy 
have been calculated for the NMSSM and  for  MSSM SPS~1a scenario at $\sqrt{s} =
500$ GeV using  CalcHEP.  Because of the admixture of a singlet in the lightest
neutralino, the cross section as a function of energy 
for the radiative neutralino production is much lower in NMSSM than in  MSSM.
We have also studied the dependence of the cross section for radiative
neutralino production on the $SU(2)_L$ gaugino mass parameter 
$M_2$ and the Higgs(ino) mass parameter $\mu$, 
as well as its dependence on the selectron~($\tilde e_R, \tilde e_L$)
masses in NMSSM, and compared it with the corresponding results in
MSSM. In order to quantify whether an excess of signal 
photons, $N_{\mathrm{S}}$, can be measured over the background
photons, $N_{\rm B}$, from radiative neutrino production, we have
analysed the theoretical statistical significance $S = N_{\rm S}/\sqrt{
N_{\rm S} + N_{\rm B}}$, and studied its dependence on $M_2$ and 
$\mu$, the parameters that enter the neutralino mass matrix.
The theoretical significance increases with the parameter $\mu$ for
both NMSSM as well as MSSM, whereas it decreases as a function of $M_2$.
The decrease is especially sharp in the case of minimal supersymmetric 
standard model. It may be interesting to study whether the signal
for radiative neutralino production in NMSSM can be enhanced by using
polarized beams~\cite{BPS}.


\section{Acknowledgements}
P.~N.~P. would like to thank The Institute of Mathematical Sciences, Chennai,
where this work was started, for its hospitality. He  would also like to thank
the Abdus Salam ICTP, where this work was completed, for its hospitality.
The work of P.~N.~P. is supported by the Board of Nuclear Sciences, Department
of Atomic Energy, India under project No. 2007/37/34/BRNS/1970.

\newpage

\end{document}